\documentclass{aa}  
\usepackage[varg]{txfonts}
\usepackage{longtable}
\usepackage{placeins}
\usepackage[inline]{enumitem}
\usepackage{comment}
\usepackage{float}
\usepackage{soul}
\usepackage{graphicx}

\usepackage{txfonts}
\usepackage[slantedGreek]{newtxmath}
\usepackage{multirow}

\usepackage{aas_macros}
\DeclareUnicodeCharacter{2212}{-}

\usepackage[dvipsnames,table]{xcolor}

\usepackage[hidelinks,colorlinks=true,allcolors=blue]{hyperref}
\usepackage{natbib}
\bibpunct{(}{)}{;}{a}{}{,} 

\usepackage[normalem]{ulem} 
\usepackage{xifthen}
\usepackage{xspace}

\def\ie{i.e.}
\def\vmax{\ensuremath{V_{\rm max}}}
\def\rmax{\ensuremath{R_{\rm max}}}
\def\mvir{\ensuremath{M_{200}}}
\def\rvir{\ensuremath{R_{200}}}
\def\frac#1#2{{\textstyle{#1\over #2}}}

\def\simlt{\stackrel{<}{{}_\sim}}

\newcommand{\LCDM}{{\normalfont$\Lambdaup$CDM}}
\newcommand{\code}[1]{\texttt{\detokenize{#1}}}
\newcommand{\cactus}{\code{CaCTus}}
\newcommand{\nexus}{{\sc nexus+}}
\newcommand{\Nbody}{{\textit{N}--body}}
\newcommand{\MillII}{{\sc millennium~II}}
\defcitealias{Hellwing2021}{H21}

\newcommand{\unit}[1]{\ensuremath{\mathrm{\,#1}}\xspace}
\newcommand{\unitlogicspace}[2]{%
  \ifthenelse{\isempty{#1}}%
    {\unit{#2}}
    {\ensuremath{{{#1}\, \unit{#2}}}}
  }

\newcommand{\Msun}[1][]{\unitlogicspace{#1}{\mathit{h}^{-1}\, M_\odot}}
\newcommand{\Mpc}[1][]{\unitlogicspace{#1}{\mathit{h}^{-1}\, Mpc}}

\newcommand{\kms}[1][]{\unitlogicspace{#1}{km\, s^{-1}}}
\usepackage{setspace}
\def\Dequal{\ensuremath{D}}
\def\Dcm{\ensuremath{\mathscr{D}}}
\def\mutrans{\ensuremath{\mu_{\rm t}}}

\def\eg{e.g.}

\begin{document} 
     \title{Caught in the cosmic web: environmental effects\\ on subhalo abundance and internal density profiles}
   \titlerunning{Environmental effects on subhalo abundance and internal density profiles}
   
   \author{Feven Markos Hunde\inst{1}
          \and
          Oliver Newton\inst{1}
          \and
          Wojciech A.~Hellwing\inst{1}
          \and
          Maciej Bilicki\inst{1}
          \and
          Krishna Naidoo\inst{2}
          }
    \institute{Center for Theoretical Physics, Polish Academy of Sciences, Al. Lotnik\'ow 32/46, 02-668 Warsaw, Poland\thanks{\email{(fevenm,onewton,hellwing)@cft.edu.pl}}
    \and
    University College London, Gower Street, London, WC1E 6BT, United Kingdom}
   \date{\today}

  \abstract{
Using the high-resolution \textit{N}--body cosmological simulation COLOR, we explore the cosmic web (CW) environmental effects on subhalo populations and their internal properties. We use \code{CaCTus}, which incorporates an implementation of the state-of-the-art segmentation method {\sc {nexus+}}, to delineate the simulation volume into nodes, filaments, walls, and voids. We group host haloes by virial mass and segment each mass bin into consecutive CW elements. This reveals that subhalo populations in hosts within specific environments differ on average from the cosmic mean. The subhalo mass function is affected strongly, where hosts in filaments typically contain more subhaloes (5 to 20\%), while hosts in voids are subhalo-poor, with 25\% fewer subhaloes. We find that the abundance of the most massive subhaloes, with reduced masses of $\mu\equiv M_\mathrm{sub}/M_{200}$ is most sensitive to the CW environment. A corresponding picture emerges when looking at subhalo mass fractions, $f_\mathrm{sub}$, where the filament hosts are significantly more `granular' (having higher $f_\mathrm{sub}$) than the cosmic mean, while the void hosts have much smoother density distributions (with $f_\mathrm{sub}$ lower by $2$ -- $20\%$ than the mean). Finally, when we look at the subhalo internal kinematic $V_{\rm max}$--$R_{\rm max}$ relations, we find that subhaloes located in the void and wall hosts exhibit density profiles with lower concentrations than the mean, while the filament hosts demonstrate much more concentrated mass profiles. Across all our samples, the effect of the CW environment generally strengthens with decreasing host halo virial mass. Our results show that host location in the large-scale CW introduces significant systematic effects on internal subhalo properties and population statistics. Understanding and accounting for them is crucial for the unbiased interpretation of observations related to small scales and satellite galaxies.
} 

   \maketitle
\section{\label{sec:intro} Introduction}

Observational efforts to map the large-scale spatial distribution of galaxies
show that they are arranged in an interconnected, filamentary network
\citep[e.g.][]{1986ApJ...302L...1D,2003astro.ph..6581C,2004ApJ...606..702T}.
This vast, complex, and multi-scale structure spans spatial scales from a few to several hundred Megaparsecs, and is now known as the `cosmic web' \citep{Bond1996}.
Numerical simulations within the framework of the $\Lambdaup+$cold dark matter~(\LCDM{}) paradigm show that the cosmic web originates from the gravitational amplification of the small Gaussian density fluctuations present within the primordial plasma \citep{Doroshkevich1970, Zeldovich1970, Shandarin1989}. Therefore, the cosmic web is one of the most significant outcomes of the anisotropic gravitational collapse governing the evolution of structure throughout the Universe \citep{1980lssu.book.....P,1981grf..book.....A,1986ApJ...304...15B}.

Classically, the cosmic web is delineated into four distinct components: high-density nodes, inhabited primarily by galaxy clusters, groups and extremely massive galaxies; elongated filaments of galaxies and intergalactic gas; flattened walls spanning the nearly empty space between the filaments; and very underdense voids \citep{2007A&A...474..315A, 2011IJMPS...1...41V, Cautun2013, Cautun2014}. To first order, this structure emerges in a well-ordered sequence as matter is evacuated from voids through walls and onto filaments, through which it is channelled towards the highest density nodes \citep{Doroshkevich1970, Cautun2014}. 

The anisotropic gravitational collapse driving the emergence of the cosmic web is also responsible for the hierarchical assembly of dark matter~(DM) haloes on smaller scales. These gravitationally bound systems grow via the merger and coalescence of smaller haloes and the smooth accretion of mass \citep{White1978,1984Natur.311..517B, White1991, 2012AnP...524..507F}. They are an essential ingredient of contemporary galaxy formation models, which posit that baryons accumulate deep within the potential wells of DM haloes and seed the growth of galaxies \citep[e.g.][]{White1978, White1991}. Thus, the properties of galaxies are sensitive to those of their host, and understanding how they are influenced by their environment is critically important.

The first attempts to connect DM halo properties with the cosmic web were carried out using gravity-only numerical simulations \citep{10.1111/j.1745-3933.2005.00084.x, Wang2007}. They show that the most massive DM haloes are typically found only in the highest density cosmic web environments \citep{Alonso2015, Hellwing2021}, and their abundance \citep{Metuki2015, Metuki2016}, concentrations \citep{2005ApJ...634...51A}, and assembly histories are similarly correlated with environmental density \citep{2007MNRAS.375..489H,10.1093/mnras/stz552}. Other works have shown that halo spins and shapes typically align with the principal axes of the filaments they are embedded within \citep{2007ApJ...655L...5A, 2007MNRAS.381...41H, 2007MNRAS.375..489H, 10.1111/j.1745-3933.2012.01222.x, 10.1093/mnras/stu1150, GVeena2018, GVeena2019, GVeena2021}, or are aligned towards the centres of the nearest void \citep{Patiri2006}. Hydrodynamic simulations show that galaxy properties such as star formation rates \citep{2019OJAp....2E...7A, Malavasi2022}, metallicity \citep{Xu2020}, and angular momentum \citep{Danovich2015} exhibit a similar dependence on the large-scale environment. Observational studies of galaxies have also shown that their mass and luminosity \citep{Wang2017}, formation times \citep{10.1093/mnras/stab2487}, morphological properties \citep{Dressler1980}, as well as the activity of the active galactic nucleus \citep{2020AJ....160..227M}, are influenced by the nature of their environment.

On even smaller scales, the correlation between the internal properties of low-mass haloes and their cosmic web environment appears to be even stronger \citep{BLlambay2013}. \cite{Hellwing2021}, hereafter \citetalias{Hellwing2021}, showed that haloes with masses below $M=\Msun[{6\times10^{10}}]$ residing in higher-density environments, such as filaments, are more concentrated compared to those found in e.g. voids. They also showed that low-mass haloes in dense environments form earlier than their counterparts with similar masses in low-density cosmic web environments. Most of the haloes in this mass regime reside within the orbit of a more massive host and were accreted during its hierarchical assembly. 

The subset of low-mass haloes that survive the tidal interactions with their hosts persist for long periods as subhaloes, some of which may host satellite galaxies \citep{1999ApJ...522...82K, 10.1046/j.1365-8711.2002.05891.x, 10.1111/j.1365-2966.2012.22111.x}. Satellite galaxies form in the highly non-linear regime of structure formation, characterised by short time scales, high densities, and chaotic orbits. They are therefore sensitive probes of the underlying cosmological model and the faint end of galaxy formation
\citep[e.g.][]{Enzi2021, Lovell2021, Newton2021, Deason2022, Dome2023, Wang2023, 2024arXiv240816042N}, and efforts to understand how the cosmic environment influences their properties is a subject of growing interest \citep{Libeskind2015,10.1093/mnras/stv919}.
Some of the best-characterised systems of low-mass galaxies currently available are the satellite populations of the Milky~Way, M31, and Centaurus~A, and they show evidence of the influence of the large-scale cosmic web environment. For example, the flattened, planar configuration of the satellites around their hosts \citep{Pawlowski2012, 2013MNRAS.431.3543H, Mueller2018} may be connected to their correlated directions of accretion via filaments \citep[][though see \citealt{PawlowskiFilamentary2012}]{Libeskind2005, Ibata2013, Cautun2015, Gillet2015, Libeskind2015, Gonzalez2016, Wang2020, Dupuy2022, GamezMarin2024}. The abundance of satellites may also be enhanced, and their specific star formation rates suppressed, compared to other similar systems if they reside in filaments \citep{Guo2015, Darvish2017}.

Over the last two decades, cosmological simulations have finally been able to reliably resolve the substructure within haloes, inciting interest in the abundance and internal properties of subhaloes in DM haloes. To characterise this information, studies have employed metrics such as the subhalo mass and velocity functions, radial distribution, and internal density profile \citep{10.1111/j.1365-2966.2004.08360.x, 10.1111/j.1365-2966.2005.08964.x, 2008MNRAS.391.1685S, 2011MNRAS.410.2309G, Gao2012, 10.1093/mnras/stu1829, 10.1093/mnras/stv2900, Hellwing2016, 10.1093/mnras/stac2930}. While previous authors have examined how these properties depend on those of their host, here we expand these analyses to consider the influence of the cosmic web environments surrounding their hosts. So far, the effect of the cosmic web on the properties of subhalo populations has not been fully explored. The goal of this paper is to address this gap in understanding and, using gravity-only numerical simulations, to connect the present-day properties of subhalo populations with the cosmic web environments of their hosts. Establishing the influence of the cosmic web environment on these properties will allow for the proper incorporation of these environmental effects into the modelling of both DM haloes and galaxy-based observables.

Using a high-resolution \Nbody{} simulation \citep{Hellwing2016} with a new implementation of the state-of-the-art multi-scale cosmic web identifier, \nexus{} \citep{Cautun2013}, we perform a systematic study of correlations between the cosmic web environment and the properties of subhalo populations. We investigate how the environment of the host haloes affects the subhalo mass and velocity functions, substructure mass fractions, and concentrations.
We find significant and robust evidence for the impact of the cosmic web environment on several subhalo population statistics.

This paper is structured as follows. In Sec.~\ref{sec:data}, we introduce the details of the COpernicus complexio LOw Resolution (COLOR) simulations and the methods that we use to ﬁnd and characterise the DM halo and subhalo populations. In Sec.~\ref{sec:cosmicweb}, we present the cosmic web segmentation algorithm used for this study. In Sec.~\ref{sec:results}, we compare different properties of DM subhaloes as a function of their locations within the cosmic web environment. In Sec.~\ref{sec:summary}, we discuss and summarise our findings and their implications, and draw our conclusion.

\section{\label{sec:data} Simulations}
In this work, we use the COpernicus complexio LOw Resolution (COLOR) simulation, which is the parent box of COpernicus COmplexio (COCO), a DM-only zoom-in \Nbody{} simulation carried out at higher resolution \citep{Bose2016, Hellwing2016}. As we aim to extend the analysis of \citetalias{Hellwing2021}
into a smaller mass regime, using the COCO CDM simulation would be a natural choice given its superb resolution.

However, as COCO is a zoom-in simulation its high-resolution volume is quite small. This restricts the size of the statistical sample that can be achieved in each cosmic web environment, especially so within void environments, which dominate the volume fraction of the cosmic web. To remedy this we choose an unusual trade-off, namely, we use the simulation with a larger volume to obtain better statistics, rather than the higher-resolution simulation that would provide better accuracy. The volume of the COLOR simulation is $14$ times larger than COCO, which significantly improves our ability to study the subhalo populations.

COLOR follows the evolution of ${\sim}4.25$~billion particles~$\left(1620^3\right)$, each with mass, $m_{\rm p} = \Msun[{6.19 \times 10^6}]$, in a periodic box with a side length of $L=\Mpc[70.4]$, encompassing a total volume of $3.5 \times 10^5 h^{-3} {\rm Mpc}^{3}$. COLOR assumes cosmological parameters obtained from the 7-year Wilkinson Microwave Anisotropy Probe (WMAP-7) data \citep{2011ApJS..192...18K}: $\Omegaup_{\rm m} = 0.272, \Omegaup_{\Lambdaup}$ = 0.728, $\Omegaup_{\rm b}$ = 0.04455, \textit{h} = 0.704, $\sigma_8$ = 0.81 and $n_{\rm s} = 0.967$. 

DM haloes were identified using the Friends-of-Friends~(FOF) algorithm with a linking length of $b = 0.2$ times the mean inter-particle separation \citep{Davis1985}. For the purposes of this paper, only haloes massive enough to host a resolved subhalo population are useful. Therefore, for our initial analysis, we consider only the FOF groups with at least $10^3$ particles, which imposes an effective mass threshold of $M_{\rm FOF}\gtrsim \Msun[10^9]$.

For each FOF group we also compute its mass, \mvir{}. This is the mass contained within the radius, \rvir, which encloses an average density of $200$ times the critical density of the Universe, $\rho_{\rm c}\!\left(z\right) \equiv 3H\! (z)^2\! / 8\pi \mathrm{G}$, at the redshift at which the halo is identified. In this analysis, we use \mvir{} as the host halo mass definition unless stated otherwise. We found gravitationally bound substructures in each FOF group using the {\sc SubFind} algorithm \citep{10.1046/j.1365-8711.2001.04912.x}, which identifies candidate subhaloes by searching for overdense regions within the FOF groups and selecting only gravitationally self-bound subhaloes. {\sc SubFind} defines the mass of a subhalo to be the total mass of all particles gravitationally bound to the overdensity. This identification process is sensitive to the resolution of the simulation. Better particle resolution improves the ability to resolve subhaloes and reduces numerical artefacts, thereby enhancing the accuracy of subhalo abundance estimates and internal density profiles \citep[\eg][]{2008MNRAS.391.1685S, Reed2005}. To minimise the effect of numerical artefacts, we require all the subhaloes we study to be composed of at least $100$ particles. We focus our analysis on the subhaloes located within the virial radius, \rvir, of their parent host halo.

Beyond resolution effects, the choice of halo and subhalo finder, as well as the method used to segment the cosmic web, can introduce systematic variations in how structures are identified and assigned to cosmic web environments. Although we do not explore alternative methods directly in this work, previous studies \citep[\eg][]{10.1111/j.1365-2966.2009.14885.x, Cautun2013, Libeskind2018} show that major cosmic web features are recovered robustly across different cosmic web classifiers, and that commonly used halo-finding algorithms typically produce consistent halo catalogues above the resolution threshold. Nevertheless, discrepancies remain in details such as subhalo boundaries and membership, particularly for low-mass or satellite systems. Based on these studies, we estimate that such methodological differences may induce shifts in the subhalo statistics by $10-20\%$ \citep[see more in][]{Knebe2013,10.1111/j.1365-2966.2012.20947.x}. We acknowledge this limitation and expect that our main conclusions, which are based on relative comparisons between environments and using conservative selection cuts (\ie~ requiring at least $100$ particles per subhalo), remain robust.

\section{\label{sec:cosmicweb} Cosmic web segmentation}
Choosing a useful physical definition to differentiate between the four cosmic web environments is not straightforward. Unlike the assignment of halo boundaries, where common choices often relate to the physical process of virialization, there is currently no simple physical criterion that allows one to formulate a single, clear definition of the cosmic web components. Due to the intrinsic multi-scale and multi-dimensional nature of the cosmic web, one must use somewhat arbitrary technical descriptions that may be less physically motivated. Popular choices include methods based on the eigenvalues of the Hessian of: the cosmic density field \citep{2007A&A...474..315A, 2007IAUS..235..127N, 10.1111/j.1365-2966.2010.17307.x}; the velocity shear tensor \citep{10.1111/j.1365-2966.2012.21553.x, 2017ApJ...845...55P}; a combination of both density field and velocity shear tensor information \citep{Cautun2013}; and the tidal or deformation tensor \citep{10.1111/j.1365-2966.2009.14885.x}. Other methods use watershed segmentations of the density field \citep{2010ApJ...723..364A}, the cosmic web skeleton construction using Morse theory \citep{10.1111/j.1365-2966.2011.18394.x}, and the identification of caustics \citep{2018JCAP...05..027F}, among others.
We refer the reader to \cite{Libeskind2018} for the most recent comprehensive presentation and comparison of the most commonly used methods.

In this work, we follow the methodology adopted by \citetalias{Hellwing2021} and use the \nexus{} approach for the cosmic web segmentation (first developed by \citealp{Cautun2013}).  In contrast to \citetalias{Hellwing2021}, we employ a new implementation of an adapted version of the \nexus{} algorithm, which will be provided in the code package \cactus{} and described in detail in \textcolor{blue}{Naidoo et al.~(in prep.)}.\footnote{\cactus{} is being prepared for public release. At the time of writing it is not yet publicly available.}
The main features of \nexus{} relevant to our study are its ability to segment the cosmic web in a multi-scale fashion, and its parametric approach that is mostly user-free and enables a self-consistent identification of the cosmic web across all scales resolved in the simulation. As the starting point we use high-resolution density fields calculated for the COLOR simulations using a simple Cloud-In-Cell~(CIC) technique. The DM density field is computed on a cubic lattice with dimensions of $256^3$, with a one-dimensional grid spacing of \Mpc[0.275]. 

The classification of haloes into cosmic web environments can vary depending on the segmentation method, particularly for haloes near environmental boundaries. However, the \nexus{} method (implemented here via the \cactus{} code) offers a significant advantage over traditional Hessian-based classifiers. Unlike single-scale approaches that rely on fixed eigenvalue thresholds, \nexus{} uses a multiscale filtering strategy and self-adaptive thresholds to extract the dominant environmental signal at each location. This produces highly stable and physically motivated classifications that are less sensitive to noise or resolution effects \citep{Cautun2013, Cautun2014}.

Below we summerise the basic set-up and all essential components of the \cactus{} implementation of the \nexus{} algorithm, which consists of five steps. 
\begin{figure}
    \centering
    \includegraphics[width=\columnwidth]{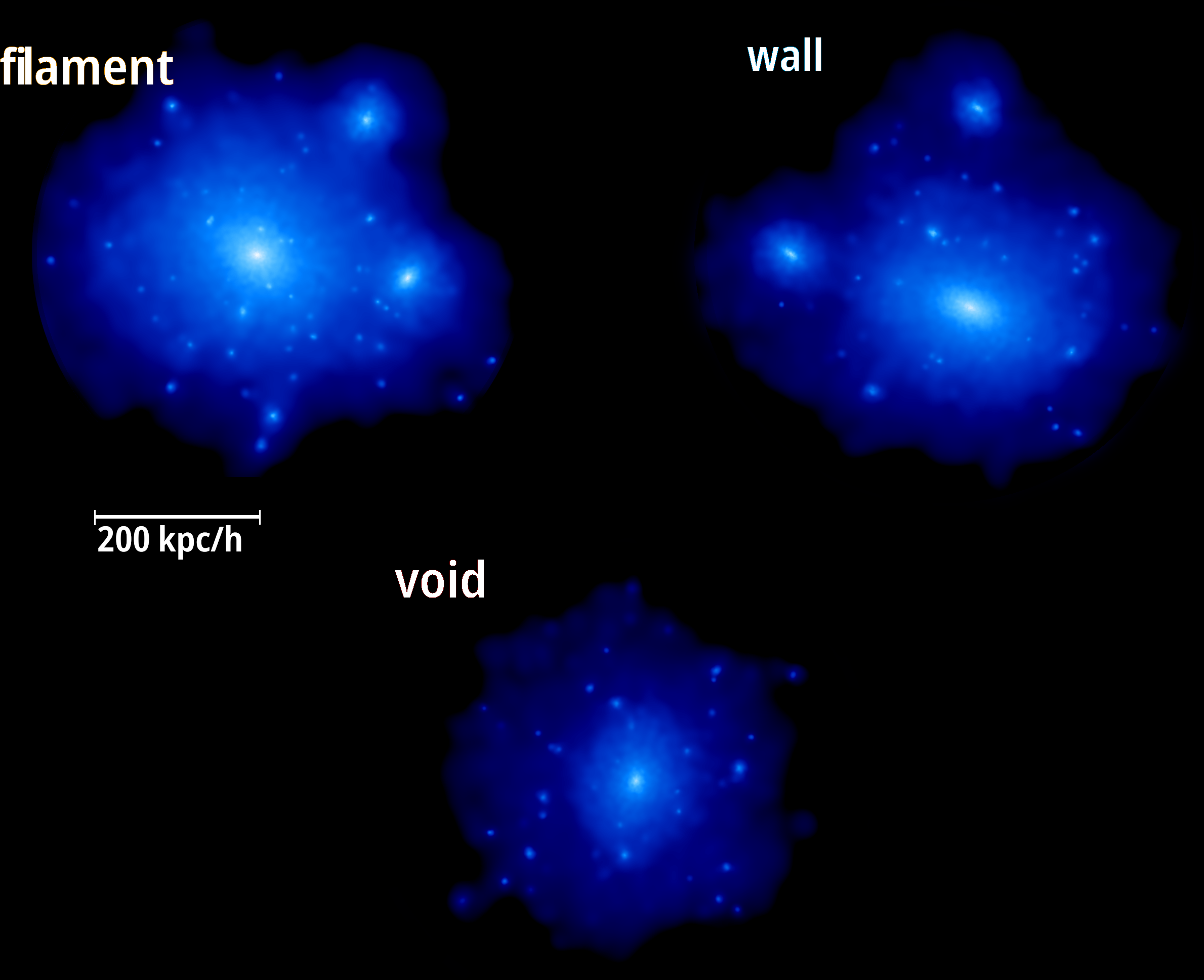}
    \caption{Renderings of the projected density of DM in three similarly massive haloes selected from different cosmic web environments identified by \cactus{}.
    Moving clockwise from the upper-left, the haloes are selected from a filament, wall, and void environment, and have masses of $\mvir{} ={8.3\times10^{11}},\, {7.0\times 10^{11}}$, and \Msun[{3.5\times 10^{11}}], respectively.
    The void halo presented here is one of the most massive found in that environment in the COLOR volume. After adjusting for the factor of two in halo mass, the most striking difference is the relative lack of massive subhaloes in the void host when compared to the other two.}
   \label{fig:hosts_cweb}
\end{figure}

\begin{table*}
     \centering
    \caption{The number density of host haloes in each host mass bin in each cosmic web environment.} 
    \setlength{\tabcolsep}{1\tabcolsep}
    \label{tab:number_host_haloes}
    \renewcommand{\arraystretch}{1.3}
    \begin{tabular}{c l l l l} 
   \hline
    $\langle \mvir{}\rangle$  & \multicolumn{4}{c}{Host halo number density $\left({\rm Mpc}^{-3}h^3\right)$} \\
    $\left(\Msun{}\right)$  & \multicolumn{1}{c}{Cosmic mean}  & \multicolumn{1}{c}{Filaments}  & \multicolumn{1}{c}{Walls} & \multicolumn{1}{c}{Voids} \\
    \hline
    $10^{12}$       & $1.29 \times 10^{-4}$ (2149)  & $1.23 \times 10^{-4}$ (2047)& $6.75 \times 10^{-6}$ (130)  & ---\\
    
    $10^{11}$       & $1.01 \times 10^{-3}$ (16711)  & $5.68 \times 10^{-4}$ (8640) & $4.14 \times 10^{-4}$ (7438)& $2.50 \times 10^{-5}$ (483)  \\
    
    $10^{10}$       & $7.72 \times 10^{-3}$ (127972)   & $2.87 \times 10^{-3}$ (45885) & $3.71 \times 10^{-3}$ (62250) &  $1.13 \times 10^{-3}$  (18736) \\
    \hline
\label{tab:numhost}
    \end{tabular}
    \vspace{-10pt}
    \tablefoot{ Numbers in brackets are the total counts of host haloes. Note that we only consider haloes that have at least 1000 particles. As the void environment contains only one halo in the $\langle\mvir{}\rangle$=\Msun[10^{12}] mass bin, we do not provide any value there.}
\end{table*}

\begin{enumerate}
    
    \item First, we smooth the DM density field, 
    \begin{equation}
        f\!(\vec{x}) \equiv \frac{\rho(\vec{x})}{\langle \rho\rangle}\,,
    \end{equation}
    with a Gaussian filter and adopt several smoothing scales, $R_n,$ resulting in a smoothed field, $f_{R_{n}}\!(\vec{x})\,.$ The smoothing scales are chosen in increments of $R_n=2^{n/2}R_0\,,$ where $R_{0}=\Mpc[0.5]$ represents the smallest scale where structures are expected to be detected. For the COLOR simulation, we use seven filter scales ranging from $0.5$ to \Mpc[4]. For filament and wall environments, the signatures are computed from smoothed log-density fields, while for the nodes, the signatures are computed with usual density smoothing. This is because logarithmic smoothing helps to enhance the asymmetric web-like features of filaments and walls, but is not necessary for nodes, which are roughly spherical structures \citep[see more in][]{Cautun2014, Cautun2015}.\\
    
    \item Second, for each filtering scale, the Hessian matrix, $\mathrm{H}_{ij,R_{n}}(\vec{x})$, is calculated via 
    \begin{equation}
     {\rm H}_{ij,R_{n}}\!(\vec{x}) = R_{n}^2 \; \left(\partial^2f_{R_{n}(\vec{x})}\big{/}{\partial x_{i}\partial x_{j}}\right)\,,
    \end{equation}
    where ${\rm H}_{ij}$ is the Hessian of the smoothed density field, and $R_{n}^2$ is a factor to normalise the Hessian across different smoothing scales. Then, the eigenvalues of the smoothed density field corresponding to each smoothing scale are computed, and are given by ${\rm det}\left({\rm H}_{ij}-\lambda_a \mathrm{I}\right) = 0$, where $\mathrm{\lambda_1} \leq \mathrm{\lambda_2} \leq \mathrm{\lambda_3}\,.$\\
    
    \item Third, the computed eigenvalues and their associated signs are used to determine the environmental signature at each grid element, which is labelled as a node, filament, or wall. This process requires the determination of a threshold value for the environmental signature, $S_{R_{n}}\!(x)$. The precise calculations can be found in equations~(6)~and~(7) of \cite{Cautun2013}. Qualitatively, node environments correspond to regions with $\lambda_1 \approx \lambda_2 \approx \lambda_3 < 0$ with a minimum average density greater than $300 \rho_{\rm c}\!\left(z\right)$, and a minimum mass greater than \Msun[{5\times10^{14}}]. Filament environments have $\lambda_1 \approx \lambda_2 < 0$ and $|\lambda_3| \ll |\lambda_2| \backsimeq |\lambda_1|$, and walls have $\lambda_1 < 0$ and $|\lambda_1| \gg |\lambda_2| \backsimeq |\lambda_3|$. The regions that are not classified as nodes, filaments, or walls are defined as voids. 
    
    \item Fourth, steps one to three are repeated for each smoothing scale, and the resulting environmental signatures are combined into one parameter-free, multi-scale signature, $S\!\left(x\right) = \max\left[S_{R_{n}}\!(x)\right]$. This environmental signature traces the morphology of particular points independently of the smoothing scale, $R_n$.
    \item Finally, cosmic web environments are then assigned to voxels in sequence, starting with clusters, then filaments, and finally walls, following \citet{Cautun2013}. Cluster environments are obtained by finding the cluster signature threshold, $S_{\rm c}^{\rm thresh}$, where half of the cluster environments with a mass greater than \Msun[{5\times 10^{14}}] have a mean density greater than $300\rho_{\rm c}\!\left(z\right)$. The regions satisfying all of the following criteria are classified as clusters:
    \begin{itemize}
        \item The cluster signature, $S_{\rm c} > S_{\rm c}^{\rm thresh}$.
        \item The mass is greater than \Msun[{5\times10^{14}}].
        \item The mean density is greater than $300 \rho_{\rm c}\!\left(z\right)$.
    \end{itemize}
   
    For filaments, we first compute the mass, $M\!\left(S_{\rm f}\right)$, contained in regions of the simulations which are not classified as clusters with a filament signature value that is greater than or equal to $S_{\rm f}$. As $S_{\rm f}$ decreases, $M\!\left(S_{\rm f}\right)$ increases.
    To determine a reasonable filament signature threshold, we fit a polynomial to $\partial M\!\left(S_{\rm f}\right)^{2}/ \partial S_{f}$. The peak of this curve determines the filament signature threshold, such that 
    \begin{equation}
    S_{\rm f}^{\rm thresh} \equiv \max \left|\frac{\partial M\!\left(S_{\rm f}\right)^{2}}{\partial S_{\rm f}}\right|. 
    \end{equation}
    All regions that are not already classified as clusters for which $S_{\rm f} > S_{\rm f}^{\rm thresh}$ are then classified as filaments.
    We discard spurious filament classifications by removing filament structures with a volume greater than $10\, h^{-3}{\rm Mpc}^{3}$.
    The threshold signature value for wall environments, $S_{\rm w}^{\rm thresh}$, is determined
    in the same way as filaments substituting the wall environment signature, $S_{\rm w}$, in place of $S_{\rm f}$. Any regions not already classified as clusters or filaments for which $S_{\rm w} > S_{\rm w}^{\rm thresh}$ are classified as walls.
    All remaining unclassified regions are assigned as voids.
\end{enumerate}

We have compared the environmental maps for the COLOR density field at $z=0$ obtained using \cactus{} with the maps used in \citetalias{Hellwing2021}, which were produced using the original \nexus{} algorithm, and found very good agreement between the two.

\section{\label{sec:results} The dependence of subhalo population properties on the cosmic web}
\begin{figure*}
    \centering
    \includegraphics[width=\textwidth]{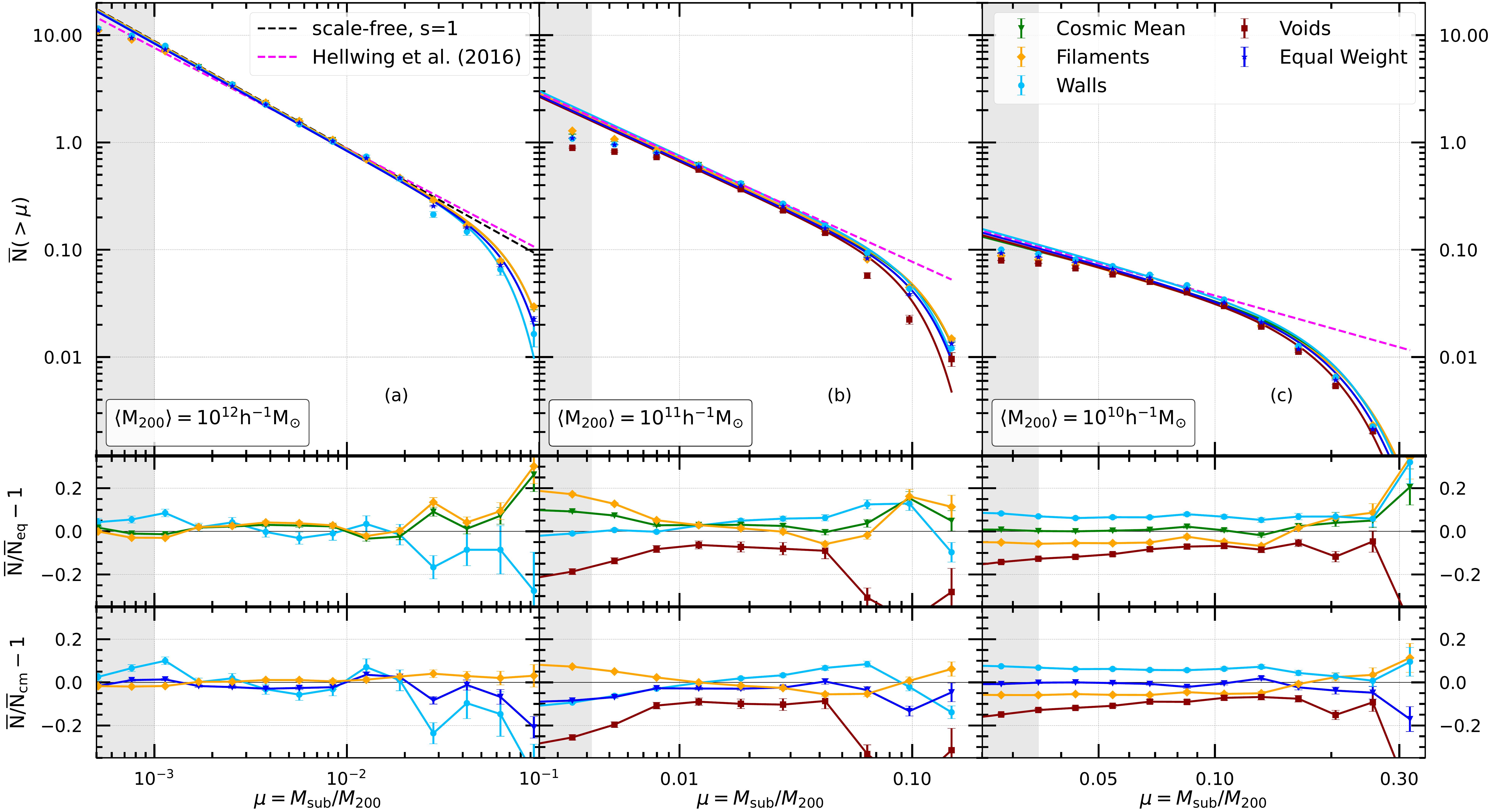}
    \caption{Upper panels: The mean cumulative subhalo mass functions for host haloes in three different mass bins, \(\langle \mvir \rangle\), shown as data points, sampled from different environments: cosmic mean (green triangles), filaments (yellow diamonds), walls (light blue circles), voids (dark red squares), and our synthetic equally weighted sample (dark blue stars). The error bars represent bootstrapped \(1\sigma\) uncertainties. The solid lines with matching colours indicate the best fits to the exponential power-law model from Eq.~(\ref{eq:fit}). The gray shaded regions on the left of each panel indicate the subhalo resolution limit, \(\mu_{\rm min}\). The two dashed lines illustrate the single power-law models for the scale-free case (black) and the best fit from the COCO simulations (magenta) of \cite{Hellwing2016}. Middle panels: The fractional deviation of each subhalo mass function with respect to the equally weighted sample, \(\Dequal\). Lower panels: The fractional deviation with respect to the cosmic mean, \(\Dcm\). The error bars in the ratio panels show uncertainties calculated through standard error propagation. Note the different ranges of the \(x\)-axes among the panels.}
   \label{fig:shmf}
\end{figure*}

In this Section, we study several properties of the subhalo populations within the context of the cosmic web environment in which the parent halo is located. First, we bin all host haloes in three ranges of virial mass corresponding to the median values of $\langle\mvir\rangle=10^{10},\, 10^{11},$ and $\Msun[10^{12}]$. Within each bin, we divide the host haloes into filaments, walls, and voids. \citetalias{Hellwing2021} showed that most haloes found in `node' environments are either central clusters, group-mass haloes, or have been strongly processed by the tidal forces in the node (e.g. the so-called backsplash haloes). 
For the former, a separate environmental analysis is not warranted because this category of hosts can be treated as clusters and galaxy group hosts. For the latter, the comparison with the rest of our sample is problematic. Most haloes in filaments, walls, and voids are field haloes, which facilitates the creation of samples of unperturbed host haloes in each environment. In contrast, the low-mass host haloes in samples constructed from node environments are often strongly processed and may be tidally disrupted. For example, the concentration--mass relation established for node haloes in \citetalias{Hellwing2021} deviates by many sigma from expectations. For these reasons, in our analysis, we ignore the `node' environment and focus only on the host halo samples in filaments, walls, and voids. As a basic illustration of how subhalo systems compare between hosts located in different cosmic web elements, we show Fig.~\ref{fig:hosts_cweb}.

Additionally, we introduce a synthetic sample called the equally weighted sample, constructed by reweighting the full population of host haloes in each cosmic web environment so that the effective contribution from each environment matches that of the environment with the smallest number of host haloes within each mass bin. This approach mitigates the sensitivity of our results to cosmic variance, which depends on the specific simulation volume we use. It also allows us to control for the bias introduced by using the cosmic mean sample, which consists of the combined set of all host haloes from walls, filaments, and voids in each mass bin. This tends to favour the richest environment, which across all mass bins is the filament environment. Throughout this paper, we characterise environmental trends by expressing subhalo properties as ratios relative to two distinct reference samples: (i) the cosmic mean, and (ii) the equally weighted sample. We denote the fractional deviation with respect to the cosmic mean as $\Dcm\equiv X/X_{\rm cm} - 1$, where $X$ denotes a given subhalo statistic or property of interest. The deviation relative to the equally weighted sample is denoted as $\Dequal\equiv X/X_{\rm eq} - 1$. These two measures, \Dcm{} and \Dequal{}, allow us to probe global environmental effects while reducing sensitivity to the specific cosmic web makeup of the simulation box.

For the range of masses of our hosts, the halo mass function is a steep power-law. Thus, constructing host mass bins in decades of fixed width would skew the halo mass distribution towards the lower-mass end of the bin. To alleviate this problem, we group the host haloes into three bins in such a way that the resulting median masses in each bin are: $\langle \mvir{}\rangle = \Msun[{\left[10^{10},10^{11},10^{12}\right]}]$. Table~\ref{tab:numhost} shows the number density of the host haloes (the number of haloes in a given cosmic web environment divided by the volume occupied by that specific environment) in each mass bin across different environments. We provide the range of host halo masses spanned by each bin in each cosmic web environment in Table ~\ref{tab:host_masses}. 

\begin{table}[h]
{
    \caption{The host halo mass ranges of each mass bin, $\langle \mvir{}\rangle$, for the halo sample in each cosmic web environment.}
    \label{tab:host_masses}
    \centering
    \setlength{\tabcolsep}{1\tabcolsep}
    \renewcommand{\arraystretch}{1.3}
    \begin{tabular}{c c c c}
    \hline
    \multirow{2}{*}{Sample} & \multicolumn{3}{c}{Mass bin $\langle \log \mvir{}\, /\, \Msun{}\rangle$}\\
     & 12 & 11 & 10\\
    \hline
    Cosmic mean &   {$11.72{-}12.72$} &   {$10.72{-}11.72$} &   {$9.72{-}10.72$} \\
    Filaments  &   {$11.68{-}12.77$} &   {$10.67{-}11.68$} &   {$9.73{-}10.67$} \\
    Walls &   {$11.87{-}12.71$} &   {$10.75{-}11.87$} &   {$9.71{-}10.75$}  \\
    Voids & {---} &   {$10.85{-}11.95$} &   {$9.77{-}10.85$} \\
    \hline
    \end{tabular}
    }
\end{table}

\subsection{\label{subsec:SHMF} The subhalo mass function}
\begin{table*}
    \centering
    \caption{Best-fitting $\beta$ values and power-law slope $s$ for cumulative subhalo mass functions.}
        \label{tab:fitting_parameters}
    \setlength{\tabcolsep}{1\tabcolsep}
    \renewcommand{\arraystretch}{1.3}
    \begin{tabular}{c c c c c c c} 
    \hline
    { $\langle M_{200} \rangle\, \left[\Msun{}\right]$ } & {Cosmic mean}  & {Filaments} & {Walls} & {Voids} & {Equally weighted} & {Power-law slope, $s$} \\
    \hline
        $10^{12}$ &  {$129 \pm 4$} &  {$100 \pm 9$} &  {$124 \pm 8$} &  {--} &  {$94 \pm 2$}  & $0.93 \pm 0.02$ \\
        $10^{11}$ &  {$89 \pm 6$} &  {$79 \pm 5$} &  {$95 \pm 4$} &  {$89 \pm 2$} &  {$97 \pm 3$} & $0.92 \pm 0.02$ \\
        $10^{10}$ &  {$67 \pm 2$} &  {$48 \pm 1$} &  {$54 \pm 2$} &  {$60 \pm 2$} &  {$57 \pm 1$} & $0.91 \pm 0.01$ \\
    \hline
    \end{tabular}
    \tablefoot{The best-fitting $\beta$ values are obtained from fits of Eq.~\eqref{eq:fit} to the cSHMF, shown as a function of the host halo mass, in the cosmic mean, filament, wall, void, and the equally weighted samples. The final column shows the best-fitting power-law slope, $s$, which is independent of the host halo environment.}

\end{table*}

Other works have established that the normalised abundance of haloes, or the halo mass function, may depend on environment \citep[][\citetalias{Hellwing2021}]{Cautun2014, Alonso2015, Libeskind2018}. Specifically, \citetalias{Hellwing2021} showed that within \nexus{} environments, the fraction of haloes distributed among different cosmic web elements strongly depends on the halo virial mass. Cluster and group-mass haloes with ${\mvir{}\gtrsim\Msun[10^{13}]}$ almost exclusively inhabit the densest parts of the cosmic web (\ie{} the nodes). However, as the host mass decreases, less massive host haloes gradually dominate the host halo abundance in the other three environments (\ie{} filaments, walls, and voids). The \nexus{} segmentation hints at a trend towards an equipartition into those three environments at the smallest masses probed in \citetalias{Hellwing2021}, \ie{} ${\mvir{}\approx \Msun[10^{10}]}$. The variation of halo abundance across the different cosmic web environments suggests that the averaged abundance of subhaloes should also be a function of the cosmic web environment. If the trends observed in halo abundance directly translate to subhalo abundance, then the abundance of the most massive subhaloes will be a strong function of the cosmic web environment of the host. If the most massive haloes (\ie{} proto-subhaloes) live in filaments, hosts residing in voids and walls will have fewer large subhaloes. To quantify and check this, we first look at the subhalo mass function~(SHMF).

In hierarchical structure formation models like CDM, it is convenient to express the subhalo mass in relation to the mass of its host. Thus, for each subhalo, we define the reduced subhalo mass, ${\mu \equiv M_ \mathrm{sub}/\mvir{}}$, and study the cumulative SHMF~(cSHMF), $N({\geq}\mu)$, which is less susceptible to random Poissonian fluctuations than a differential SHMF. This will become useful, especially for low-mass hosts from the void and wall environments, where, as we will show later, the hosts have few subhaloes. 

In Fig.~\ref{fig:shmf} we show the mean cSHMF, $\overline{N}(\geq\mu)$, across each environment. Each panel corresponds to hosts binned by $\langle \mvir{}\rangle$, and we differentiate the cSHMF into the following samples: hosts from the entire simulation volume (green triangles), those found in filaments (orange diamonds), walls (light-blue circles), and voids (dark-red squares). In addition, as mentioned earlier, as our unbiased reference point we plot the equally weighted sample (dark-blue stars), drawn uniformly from all of those environments. In the middle sub-panels, we show the fractional deviation, $\Dequal\!\left(\geq \mu\right),$ of the cSHMFs for each environmental sample with respect to the equally weighted sample.
By construction, an environmental sample with $\Dequal=0.2$ contains 20\% more subhaloes than the equally weighted sample, while $\Dequal=−0.2$ indicates a 20\% deficit. To provide additional context, the lower sub-panels in each figure show the fractional deviation of each environment relative to the cosmic mean, denoted $\Dequal(\geq\mu)$. This serves as a cosmologically representative baseline and illustrates how the cSHMF in each environment differs from that of the entire halo population within the corresponding host mass bin.
For the $\langle \mvir{}\rangle = \Msun[10^{12}]$ mass bin, we only find one host in the voids, thus, we ignore the void sample in that bin and focus only on the comparison between hosts in filament and wall environments. The error bars represent the standard error on the mean, estimated from 1000 bootstrap resamples.
Finally, the two dashed lines illustrate a simple single power-law model for the cSHMF of the form $\overline{N}(\geq\mu)\propto \mu^{-s}$, with the scale-free case of $s=1$ shown by the black dashed line, and the COCO simulation best-fitting cases of $s=0.94,0.93$, and $0.92$ for each respective host mass bin \citep{Hellwing2016} with magenta lines. 
Comparing our data with the dashed lines, we find that a single power-law provides a good description of the cSHMF only up to a transition point $\mu \lesssim \mutrans$. Beyond this scale, the cSHMF exhibits a significant exponential suppression relative to the power-law fit. For our three host halo mass bins, we estimate $\mutrans\sim 0.03, 0.04,$ and $0.1$, with the transition occurring at larger subhalo-to-host mass ratios for lower-mass hosts. This trend partially reflects the effects of simulation mass resolution: lower-mass host haloes can only reliably resolve relatively massive subhaloes, effectively shifting the onset of the cutoff to higher $\mu$. In all cases, the deviation from the power-law exceeds 10\% and lies beyond the $1\sigma$ bootstrap uncertainty, indicating a statistically robust departure from scale-free behaviour.
This observation primarily applies to the cosmic mean and equally weighted sample. In all cases, the void cSHMFs appear to have a noticeably steeper slope than for the hosts in COCO and the COLOR equally weighted samples. The slope for the wall sample in the most massive bin is similarly steep. Across all environmental samples in all mass bins, the shape of the cSHMF in the regime of the most massive subhaloes, $\mu \geq \mutrans$, significantly deviates from a uniform power-law.

\begin{figure*}%
    \centering
    \includegraphics[width=\textwidth]{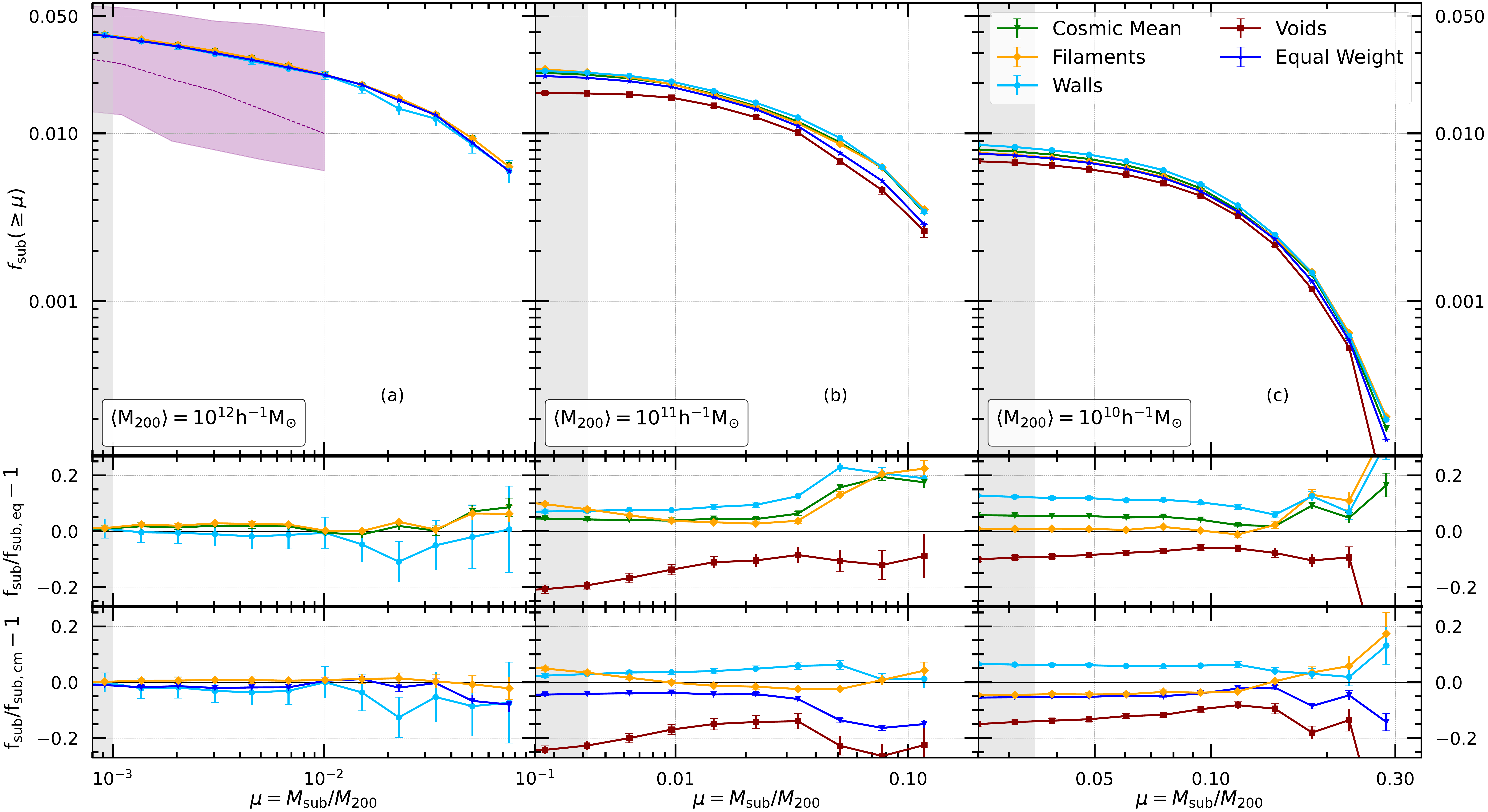}
    \caption{Upper panels: The fraction of the host halo mass in substructures. We show the same environmental samples as in Fig.~\ref{fig:shmf} for the two mass bins where the $\mu$ ranges are above the resolution limits. As before, the error bars correspond to the bootstrapped $1\sigma$ errors. The gray shaded regions at the left of each panel show the subhalo resolution limit of the simulation.
    Middle panels: The ratios of the substructure mass fraction in each environment with respect to the equally weighted sample, $\Dequal$. Lower panels: The ratios of the substructure mass fraction in each environment relative to the cosmic mean, $\Dcm$. The error bars in the ratio panel are estimated using standard error propagation. For comparison, we show the median and $60\%$ scatter (dashed line and shaded region) from the \MillII{} simulation \citep{2011MNRAS.410.2309G}.}
    \label{fig:mass_fraction}%
\end{figure*}

The exponential decay of the subhalo abundance at large $\mu$ is not surprising. Major mergers, \ie{} events that correspond to the accretion of a massive subhalo, are quite rare. This makes the capture of such subhaloes intrinsically stochastic and Poissonian \citep{Fakhouri2010}. Thus, a more realistic model for the cSHMF should take the form of an exponential power-law. To illustrate this, we consider the relationship proposed by \citet{10.1111/j.1365-2966.2010.16311.x}:
\begin{equation}
    \overline{N}(>\mu) = a\mu^{-s}\exp(-\beta \mu^3)\,,
    \label{eq:fit}
\end{equation}
where $a$ denotes the normalisation, $s$ is a power-law exponent, and $\beta$ is an exponential slope that determines the strength of the drop-off at high $\mu$ values. To determine the best-fitting values listed in Table \ref{tab:fitting_parameters}, we fit the data only down to a specific $\mu_{\mathrm{min}}$ value. For each host mass sample, $\mu_{\mathrm{min}}$ represents the minimum converged subhalo mass, which we take to be equal to $100\times m_{\rm p}$. For COLOR, this sets our limiting subhalo mass to ${M_{\rm sub}=\Msun[{6.19\times10^8}]}$. The gray shaded region represents values below the $\mu_{\mathrm{min}}$ value for each mass bin. 

\begin{table*}
    \centering
    \caption{The subhalo richness: a mean total number of subhaloes with $\mu \geq 0.04$ for all our host halo mass and environment samples.}
    \setlength{\tabcolsep}{1\tabcolsep}
    \renewcommand{\arraystretch}{1.3}
    \label{tab:subhalo_richness}
    \begin{tabular}{c c c c c c}
    \hline
    Host mass & \multicolumn{5}{c}{Subhalo richness } \\
    $\langle \mvir{}\rangle\, \left[\Msun{}\right]$ & Cosmic mean & Filaments & Walls & Voids & Equally weighted \\ \hline
    $10^{12}$ &   {$0.147 \pm 0.009$} &   {$0.147 \pm 0.009$} &   {$0.146 \pm 0.011$} & --- &   {$0.146 \pm 0.008$} \\ 
    $10^{11}$ &   {$0.138 \pm 0.003$} &   {$0.129 \pm 0.004$} &   {$0.123 \pm 0.005$} &   {$0.113 \pm 0.019$} &   {$0.124 \pm 0.013$} \\ 
    $10^{10}$ &   {$0.075 \pm 0.001$} &   {$0.074 \pm 0.001$} &   {$0.079 \pm 0.001$} &   {$0.071 \pm 0.001$} &   {$0.075 \pm 0.001$} \\ 
    \hline
    \end{tabular}
\end{table*}

Using the model from Eq.~\eqref{eq:fit}, we can now analyse our cSHMF in a more systematic way. In each cosmic web environment, the cSHMF is approximately universal within the $\mu$ regime where the SHMF closely follows a power-law, with only a weak dependence on the parent halo mass. This is in agreement with previous work \citep[][\citetalias{Hellwing2021}]{10.1111/j.1365-2966.2004.08094.x, 10.1111/j.1365-2966.2004.08360.x, 10.1111/j.1365-2966.2005.08964.x, 10.1111/j.1365-2966.2008.13182.x, 10.1111/j.1365-2966.2009.15333.x, 10.1093/mnras/stx2238,2011MNRAS.410.2309G,Jiang2016}. However, for $\mu\geq \mutrans$ the abundance of subhaloes decreases significantly and is described by a transition from a power-law to an exponential decay. The reduction of the cSHMF amplitude is especially prominent among subhaloes found in, on average, less massive haloes, which agrees with the previous results \citep{10.1111/j.1365-2966.2005.08964.x, Reed2005}.

Our results from Fig.~\ref{fig:shmf} show that the cSHMF can differ substantially between hosts with the same mass located in different cosmic web environments. We find that haloes located in filaments contain, on average, the highest number of subhaloes, with \(0.1 \leq \Dequal\left(\mu \geq \mutrans \right) < 0.2\) as shown in the middle sub-panels.
In the most massive host bin, filament subhaloes contribute the largest share of subhaloes across this $\mu$ range. For the two lower-mass host bins, subhaloes in walls outnumber those in filaments just above $\mutrans$. Filament subhaloes regain the largest contribution at the high-$\mu$ end. In comparison, perhaps unsurprisingly, hosts found in voids typically have the fewest subhaloes, with \(-0.05 \leq \Dequal\left(\mu \geq \mutrans \right) < -0.25\). In all cases, the biggest differences between environments materialise for the large $\mu$-values, \ie{} in the regime of massive and rare subhaloes. Here, at $\mu\gtrsim \mutrans$, the relative differences between the mean cSHMF of the environmental samples and the equally weighted samples, \Dequal can be as large as $0.25$. Thus, the abundance of massive subhaloes exhibits a strong dependence on the cosmic web environment of their host. This variation is captured by the best-fit values of the $\beta$-parameter, which we provide in Table~\ref{tab:fitting_parameters}.

In the low subhalo mass regime ($\mu\simlt\mutrans$), the differences between environments are more subtle but remain consistent with the trends discussed above. For the two most massive host halo bins, the cSHMF amplitude in filaments is elevated by approximately 5–20\% relative to other environments. In the lowest host mass bin, however, walls exhibit the highest subhalo abundance, exceeding other environments by 5–12\%. In contrast, voids consistently show a suppressed cSHMF that is lower by 5–25\% across all mass bins.
The cSHMFs also show that environmental effects typically become stronger with decreasing host mass. The lower sub-panels illustrate how the cSHMF in each cosmic web environment deviates from the cosmic mean, as measured by $\Dcm (\geq \mu)$. In the most massive host bin, filament subhaloes closely follow the cosmic mean, with $\Dcm\approx1$. As the host mass decreases, the cSHMF in filaments is suppressed more strongly, with \Dcm{} declining from $-0.02$ to $-0.05$. In contrast, the cSHMF in walls is enhanced as the host mass increases, with \Dcm{} rising from $0.03$ to $0.08$. The abundance of subhaloes in void hosts remains consistently suppressed across all host mass bins, and we find the largest deficit, $\Dcm = -0.15$, in massive subhaloes in the lowest-mass hosts.

Our results for the cSHMF show that the variation in subhalo abundance is both a function of the host mass and the cosmic web environment of the host. To reduce this complexity and obtain a more direct measure of the environmental effects, we look at the average expected total number of subhaloes above some mass threshold computed for hosts of different masses and cosmic web environments. We choose to count all subhaloes with $\mu\geq 0.04$ for a given host and compute the average per host mass bin and environment. We call this the subhalo richness because it corresponds to the mean expected number of subhaloes for a given host sample. We have collected the subhalo richness computed for all our host samples in Table~\ref{tab:subhalo_richness}. Subhalo richness exhibits a consistent host mass-dependent decline across environments. In the equally weighted sample, richness decreases by $\sim$14\% from the $10^{12}$ to \Msun[{10^{11}}] bin, followed by a steeper 40\% drop to the \Msun[{10^{10}}] bin. Filament and wall environments show similar trends, with initial reductions of 12\% and 16\%, respectively, between the two higher mass bins, followed by more substantial decreases of 43\% and 44\% to the lowest mass bin. In contrast to other environments, the void sample shows a more moderate reduction in subhalo richness, decreasing by 37\% from the $10^{11}$ to \Msun[{10^{10}}] bin.

The higher subhalo richness of hosts in filaments can be attributed to the dynamic environment that filaments provide for halo interactions and mergers. Haloes within filaments are situated in regions of higher matter density and experience stronger tidal forces, increasing the likelihood of substructure accretion through frequent gravitational encounters and mergers \citep{2007MNRAS.375..489H, GVeena2018}. In contrast, more underdense regions such as voids are characterised by slower structure formation, weaker tidal fields, and reduced merger activity. Consequently, haloes in voids tend to form later and accrete much less external material compared to their counterparts in denser environments \citep{Gao2004, 2007MNRAS.375..489H, Hellwing2021}.

Another interesting quantity to investigate in the context of cosmic web effects is the fraction of the total mass of haloes contained in subhaloes,
\begin{equation}
     f_\mathrm{sub}(\geq\mu_0) = \int_{\mu_0}^{1} \mu\dfrac{{\rm d}N}{{\rm d}\mu},.
     \label{eq:mass_fraction}
\end{equation}
We can understand this as an effective measure of the halo ‘granularity’. Haloes with low fractions of mass locked in subhaloes should exhibit smoother density profiles, while in contrast large $f_\mathrm{sub}$ might indicate violent virialisation in progress and/or a significant number of recent mergers. Previous studies of high-resolution \Nbody{} simulations have established that typically the mass fraction locked in subhaloes (for $\mu\gtrsim 10^{-4} {-} 10^{-2}$) ranges from $5$ to $20\%$ of the host mass \citep{10.1046/j.1365-8711.1998.01918.x, 2000ApJ...544..616G, 10.1111/j.1365-2966.2004.07372.x, 10.1111/j.1365-2966.2004.08360.x, 2011MNRAS.410.2309G, 10.1111/j.1365-2966.2011.20149.x}.

In Fig. \ref{fig:mass_fraction} we show the cumulative mass fraction in subhaloes as a function of the normalised subhalo mass, $\mu$, for our five standard samples. The middle panels display the fractional deviation of the cumulative subhalo mass fraction, $\Dequal!\left(\geq \mu\right),$ for each environmental sample relative to the equally weighted baseline. The lower sub-panels show the corresponding fractional deviation, $\Dcm!\left(\geq \mu\right),$ measured with respect to the cosmic mean sample. The variation across different cosmic web environments is not dramatic, with $f_\mathrm{sub}$ typically deviating by $2$ to $20\%$ from the equally weighted sample. For the majority of the samples, this observed effect is weak but statistically distinguishable within the $1\sigma$ bootstrap errors. There are a couple of interesting points of attention concerning the results shown in Fig.~\ref{fig:mass_fraction}. First, we note that for the $\langle M_{200} \rangle = \Msun[10^{12}]$ sample, our results agree well with what \cite{2011MNRAS.410.2309G} obtained for comparable host masses in the \textsc{Millennium-II} Simulation (shown with the dashed line and corresponding purple shaded region). For this host mass bin, the filament sample has higher mass fractions and the wall population exhibits lower values than the equally weighted sample. For the $\langle M_{200} \rangle = \Msun[10^{11}]$ bin, we also see a bimodal effect on $f_\mathrm{sub}$, but now where both filament and wall hosts have, on average, higher subhalo mass fractions, while only the void population displays lower values. The variation of $f_\mathrm{sub}$ with cosmic web environment is a sizeable effect which can potentially be important for strong lensing mass modelling studies. We leave a more thorough investigation into this matter for future studies.

\subsection{\label{subsec:SHVF} The subhalo velocity function}
\begin{figure*}
    \centering
    \includegraphics[width=\textwidth]{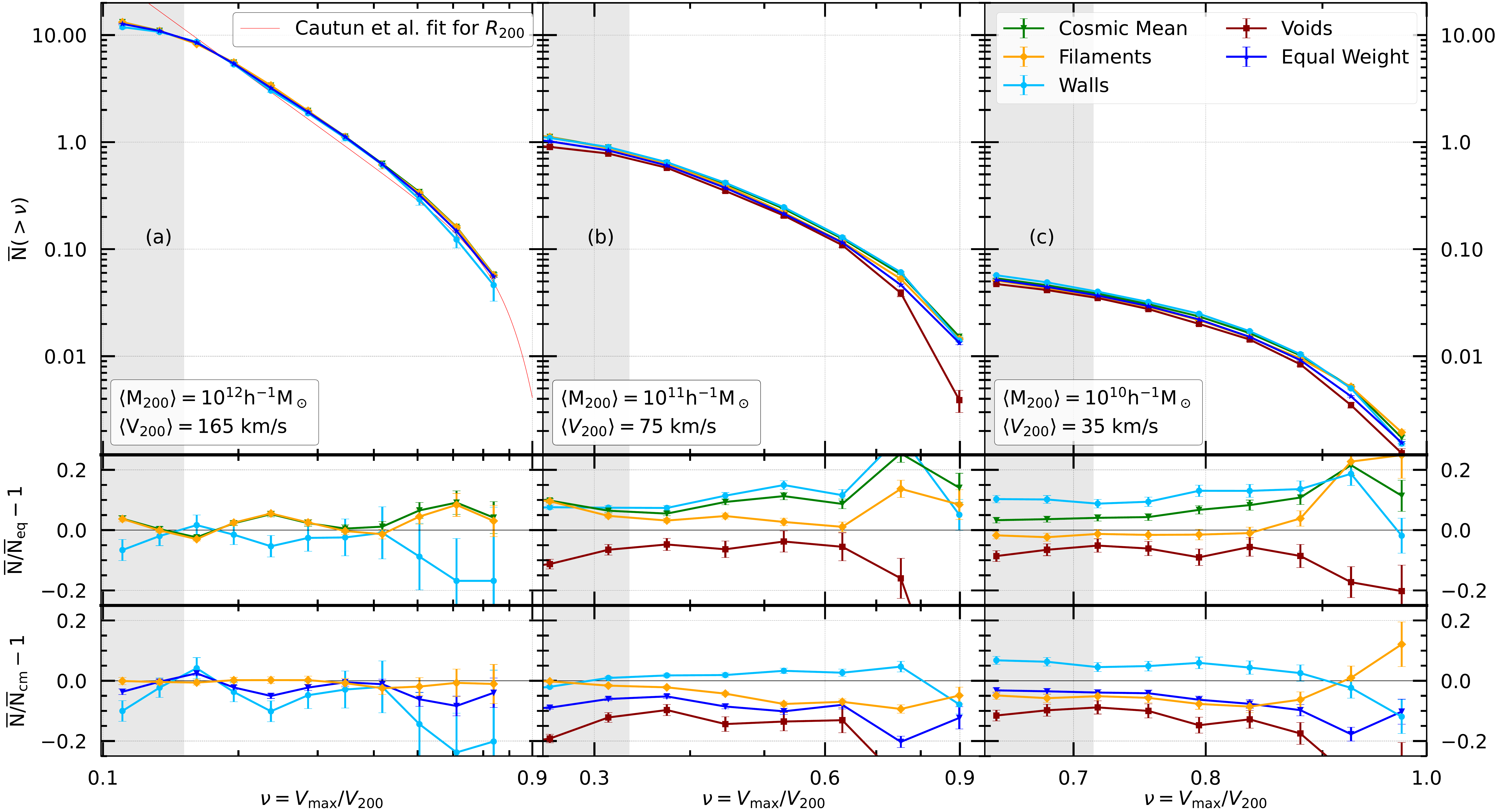}
    \caption{Upper panels: The mean cumulative distribution of subhaloes within \rvir{} as a function of $\nu=\vmax{}\, /\, V_{200}$ in the cosmic mean sample (green triangles), filaments (yellow diamonds), walls (light-blue circles), voids (dark-red squares),  and the equally weighted sample (dark-blue stars) for different host masses. The solid lines simply connect the data points, in contrast to Fig. \ref{fig:shmf} in which they represent fits to the data. In the legend, we provide the median $V_{200}$ for the given host halo mass bin. The solid red line in the figure depicts the best-fit function from \cite{10.1093/mnras/stu1829} for subhaloes within the radius \rvir{}. 
    The error bars represent the standard error on the mean, estimated from 1000 bootstrap resamples. Middle panels: The fractional deviation of $\overline{N}(>\nu)$ for each environment compared to the cumulative velocity function of an equally weighted sample, $\Dequal$. Lower panels: Deviation of $\overline{N}(>\nu)$ with respect to cosmic mean sample, $\Dcm$. Uncertainties shown in the ratio panels are computed through standard error propagation.} 
   \label{fig:shvf_1}
\end{figure*}

So far, we have used the total subhalo mass, as identified by {\sc SubFind}, to characterise the subhalo size. It is well known that, once subhaloes start orbiting within their hosts, they are subject to a strong and non-linear tidal mass stripping process. This makes the subhalo mass a less stable and sometimes even poorly defined quantity. Another way to characterise the size of a subhalo is to use the maximum of its circular orbit velocity, \vmax{}. This quantity is less affected by the strong tidal processing of the subhalo because it depends on the total enclosed mass of the subhalo. Several studies have demonstrated that for many applications \vmax{} is a robust halo size description \citep{10.1046/j.1365-8711.1998.01918.x, 2000ApJ...544..616G, 10.1111/j.1365-2966.2010.17636.x, 10.1111/j.1365-2966.2011.18858.x, 10.1111/j.1365-2966.2012.20947.x, Knebe2013, 10.1093/mnras/stu1829}.

Here, we will use the subhalo velocity functions as another measure of subhalo size and abundance. The circular velocity, denoted as $\mathrm{\textit{V}_{circ}}$, is conventionally expressed as:
\begin{equation}
   \mathrm{\textit{V}_{circ}} \equiv \sqrt{\frac{GM(<r)}{r}}
   \label{eq:vcirc}
\end{equation}
where $M\!\left(<r\right)$ is the mass enclosed inside a sphere of radius, $r$, centered at the (sub)halo centre. Each subhalo has the maximum circular velocity, \vmax{}, which represents the peak value of the circular velocity profile that is attained at a radius \rmax{}. We use \vmax{} as a stable measure of the abundance of the subhalo population. 

In Fig.~\ref{fig:shvf_1}, we show the cumulative abundance of subhaloes~(SHVF), $\overline{N}(>\nu)$, as a function of $\nu$ = $\vmax{}/{V}_{200}$, where ${V}_{200}$ is the circular velocity of the host halo at \rvir{}. The subhaloes in bins of host masses (now with the corresponding $\langle V_{200}\rangle$ values) across different environments are presented in the same manner as described in Fig. \ref{fig:shmf}. The solid red line presents the best-fit to the mean subhalo count in Milky Way-mass haloes within \rvir{} of the host from \cite{10.1093/mnras/stu1829}. The middle sub-panels show the
fractional deviation of the SHVF, $\Dequal\!\left(>\nu\right),$ of each environmental sample with respect to the equally weighted sample. The lower sub-panels present the fractional deviation compared to the cosmic mean sample, $\Dcm\!\left(>\nu\right)$. The qualitative behaviour with host samples and environments is similar to what we have seen in Fig. \ref{fig:shmf}, albeit with some slight differences. As the host halo mass decreases, the dependence on the cosmic web environment increases. The environmental dependence of the abundance of subhaloes observed in Fig.~\ref{fig:shmf} is confirmed in Fig.~\ref{fig:shvf_1}. However, while the SHMF shows a strong dependence on host halo mass, the SHVF exhibits a weaker but still noticeable dependence on host mass, consistent with findings from previous studies \citep{Hellwing2016,2023MNRAS.518..157M}. The fact that the environmental trends depicted both in SHMF and SHVF are in large qualitative agreement reassures us that the trends and effects we see here are real and robust down to the simulation resolution limit.

\subsection{The \vmax{}--\rmax{} relation}
\label{subsec:v-Rrelation}

\begin{figure*}
    \centering
    \includegraphics[width=\textwidth]{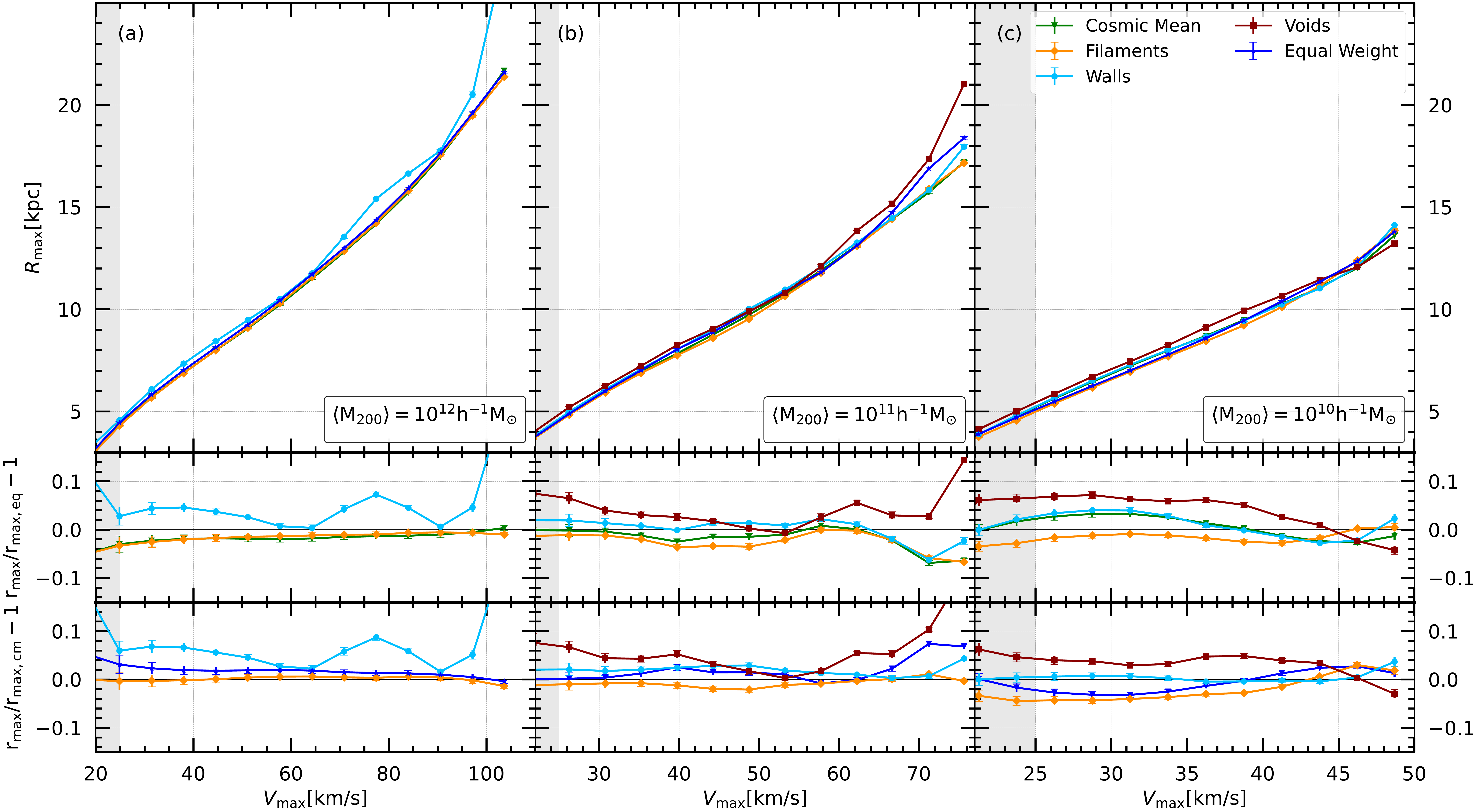}
    \caption{The mean \vmax{}--\rmax{} relation of the subhaloes within \rvir{} for different host masses across different cosmic web environments. The error bars show the bootstrap errors on the mean. The gray shaded regions at the left of each panel show the subhalo resolution limit of the simulation. Upper panels: The solid lines with error bars represent subhaloes in voids (squares), walls (circles), filaments (diamonds), the cosmic mean sample (triangles), and the equally weighted sample (stars). Middle panels: The fractional deviation of the \rmax{} values of each environment from that of the equally weighted sample, $\Dequal$. Lower panels: The relative divergence of \rmax{} values compared to the full-volume sample, $\Dcm$. In the ratio panels, the error bars represent the uncertainty obtained via standard error propagation.}

   \label{fig:v-R}
\end{figure*}

In cosmologies that describe hierarchical structure formation, such as the CDM model, halo (and thus also subhalo) density profile concentrations correlate with specific formation times. This relationship arises because, during the hierarchical assembly of (sub)haloes, the innermost regions are the first to form. Material accreted later tends to have higher angular momentum, preventing it from settling in the central parts of the (sub)halo mass distribution. Consequently, the inner regions of (sub)halo density profiles are dynamically old, and the density achieved during assembly determines the (sub)halo concentration.
As the Universe's average background density decreases over time, older (sub)haloes typically exhibit higher central densities and concentrations \citep{2002ApJ...568...52W, 10.1111/j.1365-2966.2004.08360.x, 2008MNRAS.391.1685S, 2015ApJ...810...21M, 10.1093/mnras/stw1046}. Previous works \citep[][\citetalias{Hellwing2021}]{10.1046/j.1365-8711.2001.04068.x, 10.1111/j.1365-2966.2007.11720.x, Alonso2015} have shown that halo concentration is dependent on the cosmic web environment. We want to check if this correlation is also exhibited by the host's subhalo population. This corresponds to a picture where sub(haloes) forming in dense environments are likely to be more concentrated and form earlier \citep{10.1046/j.1365-8711.2001.04068.x, 10.1111/j.1365-2966.2004.07733.x, 2005ApJ...634...51A, 10.1111/j.1365-2966.2006.10022.x, Wechsler_2006, 2007ApJ...654...53M, 10.1093/mnras/stw1046, 2020ApJ...899...81C}. 

To study the subhalo density profiles, we will exploit the \vmax{}--\rmax{} relation. This can be readily used to study the internal kinematics and the associated density profiles of (sub)haloes \citep{2008MNRAS.391.1685S, Hellwing2016}. The relation stems from a strong correlation of (sub)halo internal kinematics with the internal mass distribution, allowing for a reliable estimation of the steepness of the (sub)halo mass profile. Extra care should be taken here since the \vmax{} can be affected by the gravitational softening length ($\epsilon$). The general effect can lead to a reduction in the maximum circular velocities of subhaloes, particularly when the value of \rmax{} is not significantly greater than $\epsilon$. To address this, following \cite{Hellwing2016}, we have applied the adequate correction proposed by Eq.~(10) in \cite{2008MNRAS.391.1685S}. This correction compensates approximately for the gravitational softening effect on \vmax{}. 

In Fig.~\ref{fig:v-R}, we show the mean {\vmax{}--\rmax{}} relation for our subhalo populations in various host and environmental samples. The error bars indicate the bootstrapped uncertainty on the means. To better highlight the trends, the middle sub-panels show the fractional deviation, \Dequal, of the different samples taken with respect to the equally weighted one, while the lower sub-panels display the same quantity but computed relative to the cosmic mean sample, \Dcm. The gray shaded region represents the area below the resolution limit of the COLOR box, $\vmax{} = \kms[25]$ (see \citealt{Hellwing2016} for more detailed analysis). 

From Fig.~\ref{fig:v-R} we draw three important findings. First, the mean \rmax{} values for filament samples are lower than for other environments at fixed \vmax{}. The middle sub-panels in Fig.~\ref{fig:v-R} show the fractional deviation of subhaloes in filaments with respect to the equally weighted sample, showing \(1-8\%\) suppression. This implies that, compared to walls and voids, subhaloes found in hosts in filaments have, on average, more concentrated density profiles. Conversely, the subhalo populations residing in haloes within voids have the highest mean \rmax{} values, with showing a $1$ to $8\%$ enhancement. This suggests that subhaloes living in systems found in voids have, on average, less concentrated density profiles than the cosmic mean predicted for the whole CDM population.

Secondly, we find that the {\vmax{}--\rmax{}} relationship is affected more strongly by the environment as the host mass decreases. When examining the residuals in the $\Msun[10^{11}]$ bin (middle column of Fig.~\ref{fig:v-R}, middle sub-panel), the largest differences between subhaloes in any given environment and the equally weighted sample ranges from \(2\) to \(8\)\%. However, this increases to approximately \(3\) to \(9\)\%
in the $\Msun[10^{10}]$ bin (middle sub-panel of the right-hand column of Fig.~\ref{fig:v-R}).

The lower sub-panels of Fig.~\ref{fig:v-R} highlight how the $\rmax$--$\vmax$ relation varies with environment relative to the average across all hosts within the same mass bin. In the highest-mass bin, filament subhaloes show mean $\rmax$ values that closely follow the cosmic mean. However, toward lower host masses, their $\rmax$ values fall increasingly further below the cosmic mean, with $\Dcm$ increasing in magnitude from about 2\% to 5\% below the mean. Wall subhaloes, by contrast, start $\sim 10\%$ above the cosmic mean, then gradually approach the mean as host mass decreases. Subhaloes in voids remain consistently above the cosmic mean across all bins, with their largest deviation ($\Dcm \sim 0.08$) appearing at the low-mass end. 

One of the small-scale \LCDM{} discrepancies, which was underscored by \cite{10.1111/j.1745-3933.2011.01074.x}, is the so-called `Too Big Too Fail' (TBTF) problem. It arises from a difference between the distribution of internal properties inferred for the most massive satellites of Milky Way mass hosts in the AQUARIUS simulation suite and the internal properties of the observed dwarf Milky Way satellites \citep{10.1111/j.1365-2966.2012.20695.x}. A similar assertion was made for the field dwarf galaxies discovered in the Local Group \citep{10.1093/mnras/stu1477}. Many possible reasons have been proposed for the discrepancy, with some suggesting that processes connected to baryonic physics can lower the \vmax{} values of subhaloes hosting dwarf galaxies, while others claim that assuming a lower virial mass of the Galactic DM halo \citep{10.1093/mnras/stt259} could significantly alleviate the problem, given the strong correlation between the abundance of high \vmax{} satellites and the host virial mass \citep{10.1093/mnras/stu1849}. The second feature we observe in Fig. \ref{fig:v-R} carries significant implications for the TBTF problem.  Moving the Milky Way halo from a cosmic filament to a cosmic wall alleviates the TBTF problem by decreasing the number of high \vmax{} satellites while increasing their \rmax{} values. Our results also signify the importance of control over the cosmic environment when constructing theoretical predictions for subhalo kinematic and density distributions.

The third important observation concerns a trend associated with the mass of the parent halo. Specifically, at fixed \vmax{}, subhaloes exhibit systematically lower mean \rmax{} values when they are hosted by lower-mass haloes. To provide further detail, the environmental effects above $\vmax{} = \kms[80]$ become noticeable within the $\Msun[10^{12}]$ mass bin. The difference becomes progressively pronounced, approximately ranging from $3$ to $8\%$ across all \vmax{} values for bins within the range
$\Msun[10^{10}-10^{11}]$.

\section{\label{sec:summary}Summary and discussion}
In this paper, we have analysed the impact of the cosmic web environment on the subhalo populations of host haloes at $z=0$. To do this, we employed the \cactus{} implementation of the \nexus{} algorithm to segment the COLOR Gravity-only \Nbody{} simulation into four cosmic web environments (see sections \ref{sec:data} and \ref{sec:cosmicweb}). We disentangled the competing effects of the host halo on the subhalo properties from those of the environment by dividing the subhalo populations into bins according to the masses of their hosts. We focus on host haloes with masses in the range $\langle \mvir{} \rangle = 10^{10}$ to $\Msun[10^{12}]$. Our results extend previous studies that have already identified a strong relationship between DM halo properties and their respective locations within the cosmic web to smaller intrahalo scales.

Our main findings can be summarised as follows:
\begin{itemize}
    \item[$\bullet$] At fixed host mass, the abundance of subhaloes depends on the cosmic web environment of their hosts (Fig.~\ref{fig:shmf}). Host haloes with $\langle \mvir{} \rangle = \Msun[10^{10}]$ in walls have as much as two times more massive subhaloes than hosts with the same mass in voids. At subhalo masses below \mutrans the abundance of subhaloes in wall haloes is only $25\%$ greater than in void haloes. The size of these differences decreases as the host halo mass is increased.
    \item[$\bullet$] The measure of the halo granularity, the mass fraction in subhaloes, $f_{\text{sub}}$, shows a sizeable variation with the cosmic web (Fig.~\ref{fig:mass_fraction}). Here, typically filament hosts tend to be more granular, with 2 to 15\% more mass in subhaloes than void hosts. Void hosts, on the other hand, have 5 to 15\% less mass in subhaloes and are characterised by a smoother density distribution.
    Generally, lower mass hosts show a stronger dependence on the cosmic web.
    \item[$\bullet$] The absolute difference in velocity functions between different environments is smaller than the difference in mass functions (Fig.~\ref{fig:shvf_1}). Subhaloes in filaments have 5 to 15\% higher subhalo abundance in the velocity function, while subhaloes in voids have 5 to 25\% lower abundance relative to the cosmic mean.    
    This is because subhaloes with similar masses can have very different \vmax{}.
    \item[$\bullet$] The relative differences between the velocity functions in various environments are generally smaller than what we observe for mass functions. Generally, we start to observe greater than $10\%$ environmental difference in the velocity function above $0.5\, {V}_{200}$ for the most massive host bin. For lower-mass hosts, the differences are seen for a larger range of subhalo $\nu$ values.
    \item[$\bullet$] Subhaloes in filaments are most concentrated, and those in voids are least concentrated (Fig.~\ref{fig:v-R}). This corresponds to an approximate 5 to 9\% suppression in concentration for void subhaloes relative to filament subhaloes across all host mass bins, although in the \Msun[{10^{12}}] bin it is largely absent below $\vmax{}=\kms[80]$.
    \item[$\bullet$] The characteristic density profile also displays a dependence on the parent halo mass by showing a reduction of \rmax{} value as the host halo mass decreases. 
\end{itemize}

Our results indicate that the abundance and internal properties of subhalo populations depend on the cosmic web environment of their host. The influence of this environment becomes stronger as the host halo mass decreases below the Milky Way mass scale. Studies that do not account for this intrinsic link between the properties of subhalo populations and the large-scale environment may suffer from systematic uncertainties that bias their results.

In this paper, we analysed the SHMF as a function of $\mu = M_{\mathrm{sub}}/\mvir{}$ in different cosmic web environments. The SHMF exhibits a dependence on the cosmic web environment in which they are located. The environmental effect on the mass function becomes more noticeable with decreasing host halo mass. This is consistent with the same trend observed in the host halo mass function, as previously demonstrated in \citetalias{Hellwing2021}. Our results specifically revealed the relative differences in the SHMFs across different environments exceeding $10\%$ at lower host mass ranges. Subhaloes in voids exhibited consistently lower numbers compared to those in filaments and walls. Interestingly, the normalisation and the power-law slope of the mass functions displayed a similar pattern across all environments and with a slight dependence on host masses. However, the exponential slope, $\beta$, influencing the drop-off at the higher mass end, shows a dependence on the cosmic web environment and the parent halo mass. Furthermore, in agreement with previous studies \citep{Gao2004, 10.1111/j.1365-2966.2008.13182.x, Ishiyama_2013}, our findings demonstrate a dependency of the SHMF on the host halo mass. 

The fraction of the host mass contained in subhaloes, \( f_{\textrm{sub}} \), is highest in filament environments, indicating that filament hosts are more substructure-rich, or more `granular', than their counterparts in other cosmic web environments. This supports the idea that filaments promote subhalo accretion and survival through enhanced merger activity \citep{Bond1996, Colberg2005, Ma2025A&A.}. By contrast, haloes in voids tend to exhibit smoother density profiles, reflecting their more isolated and quiescent environments and assembly histories. These findings are consistent with the results of \citet{Hellwing2021}, who showed that halo accretion histories are affected strongly by their cosmic web environment, with filament haloes undergoing more frequent mergers.

In addition, we have analysed the subhalo velocity function to investigate an alternative perspective on the abundance of subhaloes within the cosmic web. It shows a similar dependence on the cosmic web environment like the mass function. The environmental effect on the velocity functions becomes more pronounced with decreasing parent halo mass. In a manner similar to the mass function, subhaloes hosted by haloes within filaments exhibit a higher population count compared to those in walls and voids in massive host haloes. However, a shift occurs in the case of low-mass hosts, where subhaloes hosted by haloes located in walls show a larger population than those in other environments and the relative difference is $5$ to $10\%$ higher than in the mass function. 

Furthermore, we investigated the \vmax{}--\rmax{} relation of subhaloes, which characterises the concentration or characteristic density of subhaloes based on their maximum circular velocity and radius. This comprehensive analysis revealed different density profiles among subhaloes hosted by haloes residing within filaments, walls, and voids. Particularly, subhaloes within filaments and walls exhibited, on average, more concentrated density profiles compared to those inhabiting voids. Moreover, we found that this relation is affected by the mass of their host haloes. Interestingly, a discernible trend emerged: subhalo characteristics displayed relatively lower dependence on cosmic web environments within higher-mass host haloes. However, a more substantial environmental impact became apparent with decreasing host halo mass. Notably, subhaloes hosted within filaments exhibited significantly more concentrated density profiles, particularly within lower-mass parent haloes, compared to those within massive hosts. Conversely, subhaloes in voids show on average less concentrated density profiles than in other environments. Another noteworthy feature of our results is that subhaloes tend to display lower \rmax{} values when associated with low-mass hosts, especially in contrast to subhaloes hosted by massive haloes. This finding suggests that the characteristic density of subhaloes is not exclusively determined by their environmental location but is notably influenced by the mass of the parent halo.

It is well established that the `global' properties of a subhalo, such as its mass measured at the virial radius, can be determined reliably with as few as $50{-}100$ simulation particles \citep[e.g.][]{2008MNRAS.391.1685S,10.1111/j.1365-2966.2012.20947.x,vdBoschJiang2016, Ludlow2019}. However, such low particle counts are generally insufficient to accurately characterise the internal properties of subhaloes, such as their circular velocity profiles or concentrations. Studies have shown that subhaloes resolved with fewer than $3\,000$ particles can exhibit significant deviations in their internal properties from converged results \citep{Errani2021}. Additionally, the choice of softening length and simulation time-step can further elevate the required resolution limit \citep{vdBoschOgiya2018}. The COLOR subhalo populations are not immune to numerical resolution effects, which can introduce both random and systematic biases. However, in our analysis, we mitigate these biases by studying the averaged internal properties of subhaloes within fixed mass and \vmax{} bins. Importantly, we carefully compare subhalo populations using equally weighted samples constructed from hosts within the same bin. This approach largely eliminates systematic uncertainties arising from finite resolution effects when interpreting our results. While numerical instabilities can introduce additional random scatter in the properties of subhaloes, they do not strongly affect the intrinsic differences in the average properties caused by the cosmic web environment. As a result, our findings remain robust down to the resolution limits of the simulations in the mass ranges we consider. 

An important consideration in our analysis is the relatively small volume of the COLOR simulation ($\sim100^3$ Mpc$^3$), which may introduce some level of cosmic variance in the distribution of cosmic web environments. However, comparisons of volume and mass-filling fractions between COLOR and the larger Millennium simulation \citep{2005Natur.435..629S, 2009MNRAS.398.1150B}, as reported by \citet{Hellwing2021}, indicate that such effects are unlikely to significantly impact our results. Although some variation is expected due to the finite box size, our focus on relative differences across environments-rather than absolute values-helps preserve the robustness of our conclusions. Moreover, the use of an equally weighted sampling scheme, which enforces uniform representation across environments, further mitigates biases that could arise from the overabundance of structures like filaments in the full-volume distribution. A more detailed assessment of cosmic variance, particularly in underdense environments such as voids, is deferred to future work.

Addressing the `small-scale challenges to the \LCDM{} paradigm' requires a comprehensive approach that accounts for the influence of the cosmic environment. This research is the next step toward understanding the detailed influence of large-scale structures on the distribution and formation mechanisms of satellite galaxies. We will explore the dependence of the properties of subhaloes on the cosmic web environment in alternative DM scenarios and at higher redshifts in future work.
 \begin{acknowledgements}
    We thank the anonymous referee for insightful comments and questions that have helped improve the paper. We would like to thank Marius Cautun for constructive communication and comments on the implementation of \nexus{} algorithm. Mariana Jaber, Rob Crain, Adi Nusser, Rien van de Weygaert, and Mike Hudson are acknowledged for enlightening discussions we had at various stages of this project. This work is supported via the research projects `COLAB' and `LUSTRE' funded by the National Science Center, Poland, under agreement number UMO-2020/39/B/ST9/03494 and UMO-2018/31/G/ST9/03388. MB is also supported by the Polish National Science Center through grants no. 2020/38/E/ST9/00395 and 2018/30/E/ST9/00698.
 \end{acknowledgements}

\bibliographystyle{aa_url} 
\bibliography{references}

\begin{thebibliography}{134}
\expandafter\ifx\csname natexlab\endcsname\relax\def\natexlab#1{#1}\fi

\bibitem[{{Adler}(1981)}]{1981grf..book.....A}
{Adler}, R.~J. 1981, {The Geometry of Random Fields}

\bibitem[{{Alonso} {et~al.}(2015){Alonso}, {Eardley}, \& {Peacock}}]{Alonso2015}
{Alonso}, D., {Eardley}, E., \& {Peacock}, J.~A. 2015, \href{http://dx.doi.org/10.1093/mnras/stu2632}{\color{magenta}\mnras}, \href{https://ui.adsabs.harvard.edu/abs/2015MNRAS.447.2683A}{447, 2683}

\bibitem[{Angulo {et~al.}(2009)Angulo, Lacey, Baugh, \& Frenk}]{10.1111/j.1365-2966.2009.15333.x}
Angulo, R.~E., Lacey, C.~G., Baugh, C.~M., \& Frenk, C.~S. 2009, \href{http://dx.doi.org/10.1111/j.1365-2966.2009.15333.x}{\color{magenta}\mnras}, 399, 983

\bibitem[{{Arag{\'o}n-Calvo} {et~al.}(2007{\natexlab{a}}){Arag{\'o}n-Calvo}, {Jones}, {van de Weygaert}, \& {van der Hulst}}]{2007A&A...474..315A}
{Arag{\'o}n-Calvo}, M.~A., {Jones}, B.~J.~T., {van de Weygaert}, R., \& {van der Hulst}, J.~M. 2007{\natexlab{a}}, \href{http://dx.doi.org/10.1051/0004-6361:20077880}{\color{magenta}\aap}, \href{https://ui.adsabs.harvard.edu/abs/2007A&A...474..315A}{474, 315}

\bibitem[{{Aragon Calvo} {et~al.}(2019){Aragon Calvo}, {Neyrinck}, \& {Silk}}]{2019OJAp....2E...7A}
{Aragon Calvo}, M.~A., {Neyrinck}, M.~C., \& {Silk}, J. 2019, \href{http://dx.doi.org/10.21105/astro.1697.07881}{\color{magenta}The Open Journal of Astrophysics}, \href{https://ui.adsabs.harvard.edu/abs/2019OJAp....2E...7A}{2, 7}

\bibitem[{{Arag{\'o}n-Calvo} {et~al.}(2010){Arag{\'o}n-Calvo}, {Platen}, {van de Weygaert}, \& {Szalay}}]{2010ApJ...723..364A}
{Arag{\'o}n-Calvo}, M.~A., {Platen}, E., {van de Weygaert}, R., \& {Szalay}, A.~S. 2010, \href{http://dx.doi.org/10.1088/0004-637X/723/1/364}{\color{magenta}\apj}, \href{https://ui.adsabs.harvard.edu/abs/2010ApJ...723..364A}{723, 364}

\bibitem[{{Arag{\'o}n-Calvo} {et~al.}(2007{\natexlab{b}}){Arag{\'o}n-Calvo}, {van de Weygaert}, {Jones}, \& {van der Hulst}}]{2007ApJ...655L...5A}
{Arag{\'o}n-Calvo}, M.~A., {van de Weygaert}, R., {Jones}, B. J.~T., \& {van der Hulst}, J.~M. 2007{\natexlab{b}}, \href{http://dx.doi.org/10.1086/511633}{\color{magenta}\apjl}, \href{https://ui.adsabs.harvard.edu/abs/2007ApJ...655L...5A}{655, L5}

\bibitem[{{Avila-Reese} {et~al.}(2005){Avila-Reese}, {Col{\'\i}n}, {Gottl{\"o}ber}, {Firmani}, \& {Maulbetsch}}]{2005ApJ...634...51A}
{Avila-Reese}, V., {Col{\'\i}n}, P., {Gottl{\"o}ber}, S., {Firmani}, C., \& {Maulbetsch}, C. 2005, \href{http://dx.doi.org/10.1086/491726}{\color{magenta}\apj}, \href{https://ui.adsabs.harvard.edu/abs/2005ApJ...634...51A}{634, 51}

\bibitem[{{Bardeen} {et~al.}(1986){Bardeen}, {Bond}, {Kaiser}, \& {Szalay}}]{1986ApJ...304...15B}
{Bardeen}, J.~M., {Bond}, J.~R., {Kaiser}, N., \& {Szalay}, A.~S. 1986, \href{http://dx.doi.org/10.1086/164143}{\color{magenta}\apj}, \href{https://ui.adsabs.harvard.edu/abs/1986ApJ...304...15B}{304, 15}

\bibitem[{{Ben{\'\i}tez-Llambay} {et~al.}(2013){Ben{\'\i}tez-Llambay}, {Navarro}, {Abadi}, {Gottl{\"o}ber}, {Yepes}, {Hoffman}, \& {Steinmetz}}]{BLlambay2013}
{Ben{\'\i}tez-Llambay}, A., {Navarro}, J.~F., {Abadi}, M.~G., {et~al.} 2013, \href{http://dx.doi.org/10.1088/2041-8205/763/2/L41}{\color{magenta}\apjl}, \href{https://ui.adsabs.harvard.edu/abs/2013ApJ...763L..41B}{763, L41}

\bibitem[{{Blumenthal} {et~al.}(1984){Blumenthal}, {Faber}, {Primack}, \& {Rees}}]{1984Natur.311..517B}
{Blumenthal}, G.~R., {Faber}, S.~M., {Primack}, J.~R., \& {Rees}, M.~J. 1984, \href{http://dx.doi.org/10.1038/311517a0}{\color{magenta}\nat}, \href{https://ui.adsabs.harvard.edu/abs/1984Natur.311..517B}{311, 517}

\bibitem[{{Bond} {et~al.}(1996){Bond}, {Kofman}, \& {Pogosyan}}]{Bond1996}
{Bond}, J.~R., {Kofman}, L., \& {Pogosyan}, D. 1996, \href{http://dx.doi.org/10.1038/380603a0}{\color{magenta}\nat}, \href{https://ui.adsabs.harvard.edu/abs/1996Natur.380..603B}{380, 603}

\bibitem[{Bond {et~al.}(2010)Bond, Strauss, \& Cen}]{10.1111/j.1365-2966.2010.17307.x}
Bond, N.~A., Strauss, M.~A., \& Cen, R. 2010, \href{http://dx.doi.org/10.1111/j.1365-2966.2010.17307.x}{\color{magenta}\mnras}, 409, 156

\bibitem[{Bose {et~al.}(2016)Bose, Hellwing, Frenk, Jenkins, Lovell, Helly, \& Li}]{Bose2016}
Bose, S., Hellwing, W.~A., Frenk, C.~S., {et~al.} 2016, \href{http://dx.doi.org/10.1093/mnras/stv2294}{\color{magenta}\mnras}, 455, 318

\bibitem[{Boylan-Kolchin {et~al.}(2011)Boylan-Kolchin, Bullock, \& Kaplinghat}]{10.1111/j.1745-3933.2011.01074.x}
Boylan-Kolchin, M., Bullock, J.~S., \& Kaplinghat, M. 2011, \href{http://dx.doi.org/10.1111/j.1745-3933.2011.01074.x}{\color{magenta}\mnras}, 415, L40

\bibitem[{Boylan-Kolchin {et~al.}(2012)Boylan-Kolchin, Bullock, \& Kaplinghat}]{10.1111/j.1365-2966.2012.20695.x}
Boylan-Kolchin, M., Bullock, J.~S., \& Kaplinghat, M. 2012, \href{http://dx.doi.org/10.1111/j.1365-2966.2012.20695.x}{\color{magenta}\mnras}, 422, 1203

\bibitem[{{Boylan-Kolchin} {et~al.}(2009){Boylan-Kolchin}, {Springel}, {White}, {Jenkins}, \& {Lemson}}]{2009MNRAS.398.1150B}
{Boylan-Kolchin}, M., {Springel}, V., {White}, S. D.~M., {Jenkins}, A., \& {Lemson}, G. 2009, \href{http://dx.doi.org/10.1111/j.1365-2966.2009.15191.x}{\color{magenta}\mnras}, \href{https://ui.adsabs.harvard.edu/abs/2009MNRAS.398.1150B}{398, 1150}

\bibitem[{Bullock {et~al.}(2001)Bullock, Kolatt, Sigad, Somerville, Kravtsov, Klypin, Primack, \& Dekel}]{10.1046/j.1365-8711.2001.04068.x}
Bullock, J.~S., Kolatt, T.~S., Sigad, Y., {et~al.} 2001, \href{http://dx.doi.org/10.1046/j.1365-8711.2001.04068.x}{\color{magenta}\mnras}, 321, 559

\bibitem[{{Cautun} {et~al.}(2015){Cautun}, {Bose}, {Frenk}, {Guo}, {Han}, {Hellwing}, {Sawala}, \& {Wang}}]{Cautun2015}
{Cautun}, M., {Bose}, S., {Frenk}, C.~S., {et~al.} 2015, \href{http://dx.doi.org/10.1093/mnras/stv1557}{\color{magenta}\mnras}, \href{https://ui.adsabs.harvard.edu/abs/2015MNRAS.452.3838C}{452, 3838}

\bibitem[{Cautun {et~al.}(2014{\natexlab{a}})Cautun, Frenk, van~de Weygaert, Hellwing, \& Jones}]{10.1093/mnras/stu1849}
Cautun, M., Frenk, C.~S., van~de Weygaert, R., Hellwing, W.~A., \& Jones, B. J.~T. 2014{\natexlab{a}}, \href{http://dx.doi.org/10.1093/mnras/stu1849}{\color{magenta}\mnras}, 445, 2049

\bibitem[{Cautun {et~al.}(2014{\natexlab{b}})Cautun, Hellwing, van~de Weygaert, Frenk, Jones, \& Sawala}]{10.1093/mnras/stu1829}
Cautun, M., Hellwing, W.~A., van~de Weygaert, R., {et~al.} 2014{\natexlab{b}}, \href{http://dx.doi.org/10.1093/mnras/stu1829}{\color{magenta}\mnras}, 445, 1820

\bibitem[{{Cautun} {et~al.}(2013){Cautun}, {van de Weygaert}, \& {Jones}}]{Cautun2013}
{Cautun}, M., {van de Weygaert}, R., \& {Jones}, B. J.~T. 2013, \href{http://dx.doi.org/10.1093/mnras/sts416}{\color{magenta}\mnras}, \href{https://ui.adsabs.harvard.edu/abs/2013MNRAS.429.1286C}{429, 1286}

\bibitem[{{Cautun} {et~al.}(2014){Cautun}, {van de Weygaert}, {Jones}, \& {Frenk}}]{Cautun2014}
{Cautun}, M., {van de Weygaert}, R., {Jones}, B. J.~T., \& {Frenk}, C.~S. 2014, \href{http://dx.doi.org/10.1093/mnras/stu768}{\color{magenta}\mnras}, \href{https://ui.adsabs.harvard.edu/abs/2014MNRAS.441.2923C}{441, 2923}

\bibitem[{{Chen} {et~al.}(2020){Chen}, {Mo}, {Li}, {Wang}, {Yang}, {Zhang}, \& {Wang}}]{2020ApJ...899...81C}
{Chen}, Y., {Mo}, H.~J., {Li}, C., {et~al.} 2020, \href{http://dx.doi.org/10.3847/1538-4357/aba597}{\color{magenta}\apj}, \href{https://ui.adsabs.harvard.edu/abs/2020ApJ...899...81C}{899, 81}

\bibitem[{Chua {et~al.}(2017)Chua, Pillepich, Rodriguez-Gomez, Vogelsberger, Bird, \& Hernquist}]{10.1093/mnras/stx2238}
Chua, K. T.~E., Pillepich, A., Rodriguez-Gomez, V., {et~al.} 2017, \href{http://dx.doi.org/10.1093/mnras/stx2238}{\color{magenta}\mnras}, 472, 4343

\bibitem[{{Colberg} {et~al.}(2005){Colberg}, {Krughoff}, \& {Connolly}}]{Colberg2005}
{Colberg}, J.~M., {Krughoff}, K.~S., \& {Connolly}, A.~J. 2005, \href{http://dx.doi.org/10.1111/j.1365-2966.2005.08897.x}{\color{magenta}\mnras}, \href{https://ui.adsabs.harvard.edu/abs/2005MNRAS.359..272C}{359, 272}

\bibitem[{{Colless} {et~al.}(2003){Colless}, {Peterson}, {Jackson}, {Peacock}, {Cole}, {Norberg}, {Baldry}, {Baugh}, {Bland-Hawthorn}, {Bridges}, {Cannon}, {Collins}, {Couch}, {Cross}, {Dalton}, {De Propris}, {Driver}, {Efstathiou}, {Ellis}, {Frenk}, {Glazebrook}, {Lahav}, {Lewis}, {Lumsden}, {Maddox}, {Madgwick}, {Sutherland}, \& {Taylor}}]{2003astro.ph..6581C}
{Colless}, M., {Peterson}, B.~A., {Jackson}, C., {et~al.} 2003, \href{https://ui.adsabs.harvard.edu/abs/2003astro.ph..6581C}{\href{http://dx.doi.org/10.48550/arXiv.astro-ph/0306581}{\color{magenta}arXiv e-prints}, astro}

\bibitem[{Contini {et~al.}(2012)Contini, De~Lucia, \& Borgani}]{10.1111/j.1365-2966.2011.20149.x}
Contini, E., De~Lucia, G., \& Borgani, S. 2012, \href{http://dx.doi.org/10.1111/j.1365-2966.2011.20149.x}{\color{magenta}\mnras}, 420, 2978

\bibitem[{{Danovich} {et~al.}(2015){Danovich}, {Dekel}, {Hahn}, {Ceverino}, \& {Primack}}]{Danovich2015}
{Danovich}, M., {Dekel}, A., {Hahn}, O., {Ceverino}, D., \& {Primack}, J. 2015, \href{http://dx.doi.org/10.1093/mnras/stv270}{\color{magenta}\mnras}, \href{https://ui.adsabs.harvard.edu/abs/2015MNRAS.449.2087D}{449, 2087}

\bibitem[{{Darvish} {et~al.}(2017){Darvish}, {Mobasher}, {Martin}, {Sobral}, {Scoville}, {Stroe}, {Hemmati}, \& {Kartaltepe}}]{Darvish2017}
{Darvish}, B., {Mobasher}, B., {Martin}, D.~C., {et~al.} 2017, \href{http://dx.doi.org/10.3847/1538-4357/837/1/16}{\color{magenta}\apj}, \href{https://ui.adsabs.harvard.edu/abs/2017ApJ...837...16D}{837, 16}

\bibitem[{Davis {et~al.}(1985)Davis, Efstathiou, Frenk, \& White}]{Davis1985}
Davis, M., Efstathiou, G., Frenk, C.~S., \& White, S. D.~M. 1985, \href{http://dx.doi.org/10.1086/163168}{\color{magenta}\apj}, 292

\bibitem[{{de Lapparent} {et~al.}(1986){de Lapparent}, {Geller}, \& {Huchra}}]{1986ApJ...302L...1D}
{de Lapparent}, V., {Geller}, M.~J., \& {Huchra}, J.~P. 1986, \href{http://dx.doi.org/10.1086/184625}{\color{magenta}\apjl}, \href{https://ui.adsabs.harvard.edu/abs/1986ApJ...302L...1D}{302, L1}

\bibitem[{de~Lucia {et~al.}(2004)de~Lucia, Kauffmann, Springel, White, Lanzoni, Stoehr, Tormen, \& Yoshida}]{10.1111/j.1365-2966.2004.07372.x}
de~Lucia, G., Kauffmann, G., Springel, V., {et~al.} 2004, \href{http://dx.doi.org/10.1111/j.1365-2966.2004.07372.x}{\color{magenta}\mnras}, 348, 333

\bibitem[{{Deason} {et~al.}(2022){Deason}, {Bose}, {Fattahi}, {Amorisco}, {Hellwing}, \& {Frenk}}]{Deason2022}
{Deason}, A.~J., {Bose}, S., {Fattahi}, A., {et~al.} 2022, \href{http://dx.doi.org/10.1093/mnras/stab3524}{\color{magenta}\mnras}, \href{https://ui.adsabs.harvard.edu/abs/2022MNRAS.511.4044D}{511, 4044}

\bibitem[{Diemand {et~al.}(2004)Diemand, Moore, \& Stadel}]{10.1111/j.1365-2966.2004.08094.x}
Diemand, J., Moore, B., \& Stadel, J. 2004, \href{http://dx.doi.org/10.1111/j.1365-2966.2004.08094.x}{\color{magenta}\mnras}, 353, 624

\bibitem[{{Dome} {et~al.}(2023){Dome}, {Fialkov}, {Sartorio}, \& {Mocz}}]{Dome2023}
{Dome}, T., {Fialkov}, A., {Sartorio}, N., \& {Mocz}, P. 2023, \href{http://dx.doi.org/10.1093/mnras/stad2276}{\color{magenta}\mnras}, \href{https://ui.adsabs.harvard.edu/abs/2023MNRAS.525..348D}{525, 348}

\bibitem[{{Doroshkevich}(1970)}]{Doroshkevich1970}
{Doroshkevich}, A.~G. 1970, \href{http://dx.doi.org/10.1007/BF01001625}{\color{magenta}Astrophysics}, \href{https://ui.adsabs.harvard.edu/abs/1970Ap......6..320D}{6, 320}

\bibitem[{{Dressler}(1980)}]{Dressler1980}
{Dressler}, A. 1980, \href{http://dx.doi.org/10.1086/157753}{\color{magenta}\apj}, \href{https://ui.adsabs.harvard.edu/abs/1980ApJ...236..351D}{236, 351}

\bibitem[{{Dupuy} {et~al.}(2022){Dupuy}, {Libeskind}, {Hoffman}, {Courtois}, {Gottl{\"o}ber}, {Grand}, {Knebe}, {Sorce}, {Tempel}, {Tully}, {Vogelsberger}, \& {Wang}}]{Dupuy2022}
{Dupuy}, A., {Libeskind}, N.~I., {Hoffman}, Y., {et~al.} 2022, \href{http://dx.doi.org/10.1093/mnras/stac2486}{\color{magenta}\mnras}, \href{https://ui.adsabs.harvard.edu/abs/2022MNRAS.516.4576D}{516, 4576}

\bibitem[{{Enzi} {et~al.}(2021){Enzi}, {Murgia}, {Newton}, {Vegetti}, {Frenk}, {Viel}, {Cautun}, {Fassnacht}, {Auger}, {Despali}, {McKean}, {Koopmans}, \& {Lovell}}]{Enzi2021}
{Enzi}, W., {Murgia}, R., {Newton}, O., {et~al.} 2021, \href{http://dx.doi.org/10.1093/mnras/stab1960}{\color{magenta}\mnras}, \href{https://ui.adsabs.harvard.edu/abs/2021MNRAS.506.5848E}{506, 5848}

\bibitem[{Errani \& Navarro(2021)}]{Errani2021}
Errani, R. \& Navarro, J.~F. 2021, \href{http://dx.doi.org/10.1093/mnras/stab1215}{\color{magenta}\mnras}, 505

\bibitem[{{Fakhouri} {et~al.}(2010){Fakhouri}, {Ma}, \& {Boylan-Kolchin}}]{Fakhouri2010}
{Fakhouri}, O., {Ma}, C.-P., \& {Boylan-Kolchin}, M. 2010, \href{http://dx.doi.org/10.1111/j.1365-2966.2010.16859.x}{\color{magenta}\mnras}, \href{https://ui.adsabs.harvard.edu/abs/2010MNRAS.406.2267F}{406, 2267}

\bibitem[{{Feldbrugge} {et~al.}(2018){Feldbrugge}, {van de Weygaert}, {Hidding}, \& {Feldbrugge}}]{2018JCAP...05..027F}
{Feldbrugge}, J., {van de Weygaert}, R., {Hidding}, J., \& {Feldbrugge}, J. 2018, \href{http://dx.doi.org/10.1088/1475-7516/2018/05/027}{\color{magenta}\jcap}, \href{https://ui.adsabs.harvard.edu/abs/2018JCAP...05..027F}{2018, 027}

\bibitem[{Forero-Romero {et~al.}(2014)Forero-Romero, Contreras, \& Padilla}]{10.1093/mnras/stu1150}
Forero-Romero, J.~E., Contreras, S., \& Padilla, N. 2014, \href{http://dx.doi.org/10.1093/mnras/stu1150}{\color{magenta}\mnras}, 443, 1090

\bibitem[{Forero–Romero {et~al.}(2009)Forero–Romero, Hoffman, Gottlöber, Klypin, \& Yepes}]{10.1111/j.1365-2966.2009.14885.x}
Forero–Romero, J.~E., Hoffman, Y., Gottlöber, S., Klypin, A., \& Yepes, G. 2009, \href{http://dx.doi.org/10.1111/j.1365-2966.2009.14885.x}{\color{magenta}\mnras}, 396, 1815

\bibitem[{{Frenk} \& {White}(2012)}]{2012AnP...524..507F}
{Frenk}, C.~S. \& {White}, S.~D.~M. 2012, \href{http://dx.doi.org/10.1002/andp.201200212}{\color{magenta}Annalen der Physik}, \href{https://ui.adsabs.harvard.edu/abs/2012AnP...524..507F}{524, 507}

\bibitem[{{G{\'a}mez-Mar{\'\i}n} {et~al.}(2024){G{\'a}mez-Mar{\'\i}n}, {Santos-Santos}, {Dom{\'\i}nguez-Tenreiro}, {Pedrosa}, {Tissera}, {G{\'o}mez-Flechoso}, \& {Artal}}]{GamezMarin2024}
{G{\'a}mez-Mar{\'\i}n}, M., {Santos-Santos}, I., {Dom{\'\i}nguez-Tenreiro}, R., {et~al.} 2024, \href{http://dx.doi.org/10.3847/1538-4357/ad27da}{\color{magenta}\apj}, \href{https://ui.adsabs.harvard.edu/abs/2024ApJ...965..154G}{965, 154}

\bibitem[{{Ganeshaiah Veena} {et~al.}(2019){Ganeshaiah Veena}, {Cautun}, {Tempel}, {van de Weygaert}, \& {Frenk}}]{GVeena2019}
{Ganeshaiah Veena}, P., {Cautun}, M., {Tempel}, E., {van de Weygaert}, R., \& {Frenk}, C.~S. 2019, \href{http://dx.doi.org/10.1093/mnras/stz1343}{\color{magenta}\mnras}, \href{https://ui.adsabs.harvard.edu/abs/2019MNRAS.487.1607G}{487, 1607}

\bibitem[{{Ganeshaiah Veena} {et~al.}(2021){Ganeshaiah Veena}, {Cautun}, {van de Weygaert}, {Tempel}, \& {Frenk}}]{GVeena2021}
{Ganeshaiah Veena}, P., {Cautun}, M., {van de Weygaert}, R., {Tempel}, E., \& {Frenk}, C.~S. 2021, \href{http://dx.doi.org/10.1093/mnras/stab411}{\color{magenta}\mnras}, \href{https://ui.adsabs.harvard.edu/abs/2021MNRAS.503.2280G}{503, 2280}

\bibitem[{{Ganeshaiah Veena} {et~al.}(2018){Ganeshaiah Veena}, {Cautun}, {van de Weygaert}, {Tempel}, {Jones}, {Rieder}, \& {Frenk}}]{GVeena2018}
{Ganeshaiah Veena}, P., {Cautun}, M., {van de Weygaert}, R., {et~al.} 2018, \href{http://dx.doi.org/10.1093/mnras/sty2270}{\color{magenta}\mnras}, \href{https://ui.adsabs.harvard.edu/abs/2018MNRAS.481..414G}{481, 414}

\bibitem[{{Gao} {et~al.}(2011){Gao}, {Frenk}, {Boylan-Kolchin}, {Jenkins}, {Springel}, \& {White}}]{2011MNRAS.410.2309G}
{Gao}, L., {Frenk}, C.~S., {Boylan-Kolchin}, M., {et~al.} 2011, \href{http://dx.doi.org/10.1111/j.1365-2966.2010.17601.x}{\color{magenta}\mnras}, \href{https://ui.adsabs.harvard.edu/abs/2011MNRAS.410.2309G}{410, 2309}

\bibitem[{Gao {et~al.}(2004{\natexlab{a}})Gao, Lucia, White, \& Jenkins}]{Gao2004}
Gao, L., Lucia, G.~D., White, S.~D., \& Jenkins, A. 2004{\natexlab{a}}, \href{http://dx.doi.org/10.1111/j.1365-2966.2004.08098.x}{\color{magenta}\mnras}, 352

\bibitem[{Gao {et~al.}(2012)Gao, Navarro, Frenk, Jenkins, Springel, \& White}]{Gao2012}
Gao, L., Navarro, J.~F., Frenk, C.~S., {et~al.} 2012, \href{http://dx.doi.org/10.1111/j.1365-2966.2012.21564.x}{\color{magenta}\mnras}, 425

\bibitem[{{Gao} {et~al.}(2005){Gao}, {Springel}, \& {White}}]{10.1111/j.1745-3933.2005.00084.x}
{Gao}, L., {Springel}, V., \& {White}, S. D.~M. 2005, \href{http://dx.doi.org/10.1111/j.1745-3933.2005.00084.x}{\color{magenta}\mnras}, \href{https://ui.adsabs.harvard.edu/abs/2005MNRAS.363L..66G}{363, L66}

\bibitem[{Gao {et~al.}(2004{\natexlab{b}})Gao, White, Jenkins, Stoehr, \& Springel}]{10.1111/j.1365-2966.2004.08360.x}
Gao, L., White, S. D.~M., Jenkins, A., Stoehr, F., \& Springel, V. 2004{\natexlab{b}}, \href{http://dx.doi.org/10.1111/j.1365-2966.2004.08360.x}{\color{magenta}\mnras}, 355, 819

\bibitem[{Garrison-Kimmel {et~al.}(2014)Garrison-Kimmel, Boylan-Kolchin, Bullock, \& Kirby}]{10.1093/mnras/stu1477}
Garrison-Kimmel, S., Boylan-Kolchin, M., Bullock, J.~S., \& Kirby, E.~N. 2014, \href{http://dx.doi.org/10.1093/mnras/stu1477}{\color{magenta}\mnras}, 444, 222

\bibitem[{Ghigna {et~al.}(1998)Ghigna, Moore, Governato, Lake, Quinn, \& Stadel}]{10.1046/j.1365-8711.1998.01918.x}
Ghigna, S., Moore, B., Governato, F., {et~al.} 1998, \href{http://dx.doi.org/10.1046/j.1365-8711.1998.01918.x}{\color{magenta}\mnras}, 300, 146

\bibitem[{{Ghigna} {et~al.}(2000){Ghigna}, {Moore}, {Governato}, {Lake}, {Quinn}, \& {Stadel}}]{2000ApJ...544..616G}
{Ghigna}, S., {Moore}, B., {Governato}, F., {et~al.} 2000, \href{http://dx.doi.org/10.1086/317221}{\color{magenta}\apj}, \href{https://ui.adsabs.harvard.edu/abs/2000ApJ...544..616G}{544, 616}

\bibitem[{{Gillet} {et~al.}(2015){Gillet}, {Ocvirk}, {Aubert}, {Knebe}, {Libeskind}, {Yepes}, {Gottl{\"o}ber}, \& {Hoffman}}]{Gillet2015}
{Gillet}, N., {Ocvirk}, P., {Aubert}, D., {et~al.} 2015, \href{http://dx.doi.org/10.1088/0004-637X/800/1/34}{\color{magenta}\apj}, \href{https://ui.adsabs.harvard.edu/abs/2015ApJ...800...34G}{800, 34}

\bibitem[{Giocoli {et~al.}(2010)Giocoli, Tormen, Sheth, \& van~den Bosch}]{10.1111/j.1365-2966.2010.16311.x}
Giocoli, C., Tormen, G., Sheth, R.~K., \& van~den Bosch, F.~C. 2010, \href{http://dx.doi.org/10.1111/j.1365-2966.2010.16311.x}{\color{magenta}\mnras}, 404, 502

\bibitem[{Giocoli {et~al.}(2008)Giocoli, Tormen, \& Van Den~Bosch}]{10.1111/j.1365-2966.2008.13182.x}
Giocoli, C., Tormen, G., \& Van Den~Bosch, F.~C. 2008, \href{http://dx.doi.org/10.1111/j.1365-2966.2008.13182.x}{\color{magenta}\mnras}, 386, 2135

\bibitem[{{Gonz{\'a}lez} \& {Padilla}(2016)}]{Gonzalez2016}
{Gonz{\'a}lez}, R.~E. \& {Padilla}, N.~D. 2016, \href{http://dx.doi.org/10.3847/0004-637X/829/1/58}{\color{magenta}\apj}, \href{https://ui.adsabs.harvard.edu/abs/2016ApJ...829...58G}{829, 58}

\bibitem[{{Guo} {et~al.}(2015){Guo}, {Tempel}, \& {Libeskind}}]{Guo2015}
{Guo}, Q., {Tempel}, E., \& {Libeskind}, N.~I. 2015, \href{http://dx.doi.org/10.1088/0004-637X/800/2/112}{\color{magenta}\apj}, \href{https://ui.adsabs.harvard.edu/abs/2015ApJ...800..112G}{800, 112}

\bibitem[{{Hahn} {et~al.}(2007{\natexlab{a}}){Hahn}, {Carollo}, {Porciani}, \& {Dekel}}]{2007MNRAS.381...41H}
{Hahn}, O., {Carollo}, C.~M., {Porciani}, C., \& {Dekel}, A. 2007{\natexlab{a}}, \href{http://dx.doi.org/10.1111/j.1365-2966.2007.12249.x}{\color{magenta}\mnras}, \href{https://ui.adsabs.harvard.edu/abs/2007MNRAS.381...41H}{381, 41}

\bibitem[{{Hahn} {et~al.}(2007{\natexlab{b}}){Hahn}, {Porciani}, {Carollo}, \& {Dekel}}]{2007MNRAS.375..489H}
{Hahn}, O., {Porciani}, C., {Carollo}, C.~M., \& {Dekel}, A. 2007{\natexlab{b}}, \href{http://dx.doi.org/10.1111/j.1365-2966.2006.11318.x}{\color{magenta}\mnras}, \href{https://ui.adsabs.harvard.edu/abs/2007MNRAS.375..489H}{375, 489}

\bibitem[{{Hammer} {et~al.}(2013){Hammer}, {Yang}, {Fouquet}, {Pawlowski}, {Kroupa}, {Puech}, {Flores}, \& {Wang}}]{2013MNRAS.431.3543H}
{Hammer}, F., {Yang}, Y., {Fouquet}, S., {et~al.} 2013, \href{http://dx.doi.org/10.1093/mnras/stt435}{\color{magenta}\mnras}, \href{https://ui.adsabs.harvard.edu/abs/2013MNRAS.431.3543H}{431, 3543}

\bibitem[{Han {et~al.}(2016)Han, Cole, Frenk, \& Jing}]{10.1093/mnras/stv2900}
Han, J., Cole, S., Frenk, C.~S., \& Jing, Y. 2016, \href{http://dx.doi.org/10.1093/mnras/stv2900}{\color{magenta}\mnras}, 457, 1208

\bibitem[{Han {et~al.}(2012)Han, Jing, Wang, \& Wang}]{10.1111/j.1365-2966.2012.22111.x}
Han, J., Jing, Y.~P., Wang, H., \& Wang, W. 2012, \href{http://dx.doi.org/10.1111/j.1365-2966.2012.22111.x}{\color{magenta}\mnras}, 427, 2437

\bibitem[{Harker {et~al.}(2006)Harker, Cole, Helly, Frenk, \& Jenkins}]{10.1111/j.1365-2966.2006.10022.x}
Harker, G., Cole, S., Helly, J., Frenk, C., \& Jenkins, A. 2006, \href{http://dx.doi.org/10.1111/j.1365-2966.2006.10022.x}{\color{magenta}\mnras}, 367, 1039

\bibitem[{{Hellwing} {et~al.}(2021){Hellwing}, {Cautun}, {van de Weygaert}, \& {Jones}}]{Hellwing2021}
{Hellwing}, W.~A., {Cautun}, M., {van de Weygaert}, R., \& {Jones}, B.~T. 2021, \href{http://dx.doi.org/10.1103/PhysRevD.103.063517}{\color{magenta}\prd}, \href{https://ui.adsabs.harvard.edu/abs/2021PhRvD.103f3517H}{103, 063517}

\bibitem[{{Hellwing} {et~al.}(2016){Hellwing}, {Frenk}, {Cautun}, {Bose}, {Helly}, {Jenkins}, {Sawala}, \& {Cytowski}}]{Hellwing2016}
{Hellwing}, W.~A., {Frenk}, C.~S., {Cautun}, M., {et~al.} 2016, \href{http://dx.doi.org/10.1093/mnras/stw214}{\color{magenta}\mnras}, \href{https://ui.adsabs.harvard.edu/abs/2016MNRAS.457.3492H}{457, 3492}

\bibitem[{Hoffman {et~al.}(2012)Hoffman, Metuki, Yepes, Gottlöber, Forero-Romero, Libeskind, \& Knebe}]{10.1111/j.1365-2966.2012.21553.x}
Hoffman, Y., Metuki, O., Yepes, G., {et~al.} 2012, \href{http://dx.doi.org/10.1111/j.1365-2966.2012.21553.x}{\color{magenta}\mnras}, 425, 2049

\bibitem[{{Ibata} {et~al.}(2013){Ibata}, {Lewis}, {Conn}, {Irwin}, {McConnachie}, {Chapman}, {Collins}, {Fardal}, {Ferguson}, {Ibata}, {Mackey}, {Martin}, {Navarro}, {Rich}, {Valls-Gabaud}, \& {Widrow}}]{Ibata2013}
{Ibata}, R.~A., {Lewis}, G.~F., {Conn}, A.~R., {et~al.} 2013, \href{http://dx.doi.org/10.1038/nature11717}{\color{magenta}\nat}, \href{https://ui.adsabs.harvard.edu/abs/2013Natur.493...62I}{493, 62}

\bibitem[{Ishiyama {et~al.}(2013)Ishiyama, Rieder, Makino, Portegies~Zwart, Groen, Nitadori, de~Laat, McMillan, Hiraki, \& Harfst}]{Ishiyama_2013}
Ishiyama, T., Rieder, S., Makino, J., {et~al.} 2013, \href{http://dx.doi.org/10.1088/0004-637X/767/2/146}{\color{magenta}The Astrophysical Journal}, 767, 146

\bibitem[{Jiang \& Bosch(2016)}]{Jiang2016}
Jiang, F. \& Bosch, F. C. V.~D. 2016, \href{http://dx.doi.org/10.1093/mnras/stw439}{\color{magenta}\mnras}, 458

\bibitem[{{Klypin} {et~al.}(1999){Klypin}, {Kravtsov}, {Valenzuela}, \& {Prada}}]{1999ApJ...522...82K}
{Klypin}, A., {Kravtsov}, A.~V., {Valenzuela}, O., \& {Prada}, F. 1999, \href{http://dx.doi.org/10.1086/307643}{\color{magenta}\apj}, \href{https://ui.adsabs.harvard.edu/abs/1999ApJ...522...82K}{522, 82}

\bibitem[{Knebe {et~al.}(2011)Knebe, Knollmann, Muldrew, Pearce, Aragon-Calvo, Ascasibar, Behroozi, Ceverino, Colombi, Diemand, Dolag, Falck, Fasel, Gardner, Gottlöber, Hsu, Iannuzzi, Klypin, Lukić, Maciejewski, McBride, Neyrinck, Planelles, Potter, Quilis, Rasera, Read, Ricker, Roy, Springel, Stadel, Stinson, Sutter, Turchaninov, Tweed, Yepes, \& Zemp}]{10.1111/j.1365-2966.2011.18858.x}
Knebe, A., Knollmann, S.~R., Muldrew, S.~I., {et~al.} 2011, \href{http://dx.doi.org/10.1111/j.1365-2966.2011.18858.x}{\color{magenta}\mnras}, 415, 2293

\bibitem[{Knebe {et~al.}(2013)Knebe, Pearce, Lux, Ascasibar, Behroozi, Casado, Moran, Diemand, Dolag, Dominguez-Tenreiro, Elahi, Falck, Gottlöber, Han, Klypin, Lukić, Maciejewski, McBride, Merchán, Muldrew, Neyrinck, Onions, Planelles, Potter, Quilis, Rasera, Ricker, Roy, Ruiz, Sgró, Springel, Stadel, Sutter, Tweed, \& Zemp}]{Knebe2013}
Knebe, A., Pearce, F.~R., Lux, H., {et~al.} 2013, \href{http://dx.doi.org/10.1093/mnras/stt1403}{\color{magenta}\mnras}, 435

\bibitem[{{Komatsu} {et~al.}(2011){Komatsu}, {Smith}, {Dunkley}, {Bennett}, {Gold}, {Hinshaw}, {Jarosik}, {Larson}, {Nolta}, {Page}, {Spergel}, {Halpern}, {Hill}, {Kogut}, {Limon}, {Meyer}, {Odegard}, {Tucker}, {Weiland}, {Wollack}, \& {Wright}}]{2011ApJS..192...18K}
{Komatsu}, E., {Smith}, K.~M., {Dunkley}, J., {et~al.} 2011, \href{http://dx.doi.org/10.1088/0067-0049/192/2/18}{\color{magenta}\apjs}, \href{https://ui.adsabs.harvard.edu/abs/2011ApJS..192...18K}{192, 18}

\bibitem[{{Libeskind} {et~al.}(2005){Libeskind}, {Frenk}, {Cole}, {Helly}, {Jenkins}, {Navarro}, \& {Power}}]{Libeskind2005}
{Libeskind}, N.~I., {Frenk}, C.~S., {Cole}, S., {et~al.} 2005, \href{http://dx.doi.org/10.1111/j.1365-2966.2005.09425.x}{\color{magenta}\mnras}, \href{https://ui.adsabs.harvard.edu/abs/2005MNRAS.363..146L}{363, 146}

\bibitem[{{Libeskind} {et~al.}(2012){Libeskind}, {Hoffman}, {Knebe}, {Steinmetz}, {Gottl{\"o}ber}, {Metuki}, \& {Yepes}}]{10.1111/j.1745-3933.2012.01222.x}
{Libeskind}, N.~I., {Hoffman}, Y., {Knebe}, A., {et~al.} 2012, \href{http://dx.doi.org/10.1111/j.1745-3933.2012.01222.x}{\color{magenta}\mnras}, \href{https://ui.adsabs.harvard.edu/abs/2012MNRAS.421L.137L}{421, L137}

\bibitem[{{Libeskind} {et~al.}(2015){Libeskind}, {Hoffman}, {Tully}, {Courtois}, {Pomar{\`e}de}, {Gottl{\"o}ber}, \& {Steinmetz}}]{Libeskind2015}
{Libeskind}, N.~I., {Hoffman}, Y., {Tully}, R.~B., {et~al.} 2015, \href{http://dx.doi.org/10.1093/mnras/stv1302}{\color{magenta}\mnras}, \href{https://ui.adsabs.harvard.edu/abs/2015MNRAS.452.1052L}{452, 1052}

\bibitem[{Libeskind {et~al.}(2018)Libeskind, van~de Weygaert, Cautun, Falck, Tempel, Abel, Alpaslan, Aragón-Calvo, Forero-Romero, Gonzalez, Gottlöber, Hahn, Hellwing, Hoffman, Jones, Kitaura, Knebe, Manti, Neyrinck, Nuza, Padilla, Platen, Ramachandra, Robotham, Saar, Shandarin, Steinmetz, Stoica, Sousbie, \& Yepes}]{Libeskind2018}
Libeskind, N.~I., van~de Weygaert, R., Cautun, M., {et~al.} 2018, \href{http://dx.doi.org/10.1093/mnras/stx1976}{\color{magenta}\mnras}, 473, 1195

\bibitem[{{Lovell} {et~al.}(2021){Lovell}, {Cautun}, {Frenk}, {Hellwing}, \& {Newton}}]{Lovell2021}
{Lovell}, M.~R., {Cautun}, M., {Frenk}, C.~S., {Hellwing}, W.~A., \& {Newton}, O. 2021, \href{http://dx.doi.org/10.1093/mnras/stab2452}{\color{magenta}\mnras}, \href{https://ui.adsabs.harvard.edu/abs/2021MNRAS.507.4826L}{507, 4826}

\bibitem[{Ludlow {et~al.}(2016)Ludlow, Bose, Angulo, Wang, Hellwing, Navarro, Cole, \& Frenk}]{10.1093/mnras/stw1046}
Ludlow, A.~D., Bose, S., Angulo, R.~E., {et~al.} 2016, \href{http://dx.doi.org/10.1093/mnras/stw1046}{\color{magenta}\mnras}, 460, 1214

\bibitem[{{Ludlow} {et~al.}(2019){Ludlow}, {Schaye}, \& {Bower}}]{Ludlow2019}
{Ludlow}, A.~D., {Schaye}, J., \& {Bower}, R. 2019, \href{http://dx.doi.org/10.1093/mnras/stz1821}{\color{magenta}\mnras}, \href{https://ui.adsabs.harvard.edu/abs/2019MNRAS.488.3663L}{488, 3663}

\bibitem[{{Ma} {et~al.}(2025){Ma}, {Guo}, \& {Jones}}]{Ma2025A&A.}
{Ma}, W., {Guo}, H., \& {Jones}, M.~G. 2025, \href{http://dx.doi.org/10.1051/0004-6361/202451932}{\color{magenta}\aap}, \href{https://ui.adsabs.harvard.edu/abs/2025A&A...695A...5M}{695, A5}

\bibitem[{Macciò {et~al.}(2007)Macciò, Dutton, Van Den~Bosch, Moore, Potter, \& Stadel}]{10.1111/j.1365-2966.2007.11720.x}
Macciò, A.~V., Dutton, A.~A., Van Den~Bosch, F.~C., {et~al.} 2007, \href{http://dx.doi.org/10.1111/j.1365-2966.2007.11720.x}{\color{magenta}\mnras}, 378, 55

\bibitem[{{Malavasi} {et~al.}(2022){Malavasi}, {Langer}, {Aghanim}, {Gal{\'a}rraga-Espinosa}, \& {Gouin}}]{Malavasi2022}
{Malavasi}, N., {Langer}, M., {Aghanim}, N., {Gal{\'a}rraga-Espinosa}, D., \& {Gouin}, C. 2022, \href{http://dx.doi.org/10.1051/0004-6361/202141723}{\color{magenta}\aap}, \href{https://ui.adsabs.harvard.edu/abs/2022A&A...658A.113M}{658, A113}

\bibitem[{{Mao} {et~al.}(2015){Mao}, {Williamson}, \& {Wechsler}}]{2015ApJ...810...21M}
{Mao}, Y.-Y., {Williamson}, M., \& {Wechsler}, R.~H. 2015, \href{http://dx.doi.org/10.1088/0004-637X/810/1/21}{\color{magenta}\apj}, \href{https://ui.adsabs.harvard.edu/abs/2015ApJ...810...21M}{810, 21}

\bibitem[{{Maulbetsch} {et~al.}(2007){Maulbetsch}, {Avila-Reese}, {Col{\'\i}n}, {Gottl{\"o}ber}, {Khalatyan}, \& {Steinmetz}}]{2007ApJ...654...53M}
{Maulbetsch}, C., {Avila-Reese}, V., {Col{\'\i}n}, P., {et~al.} 2007, \href{http://dx.doi.org/10.1086/509706}{\color{magenta}\apj}, \href{https://ui.adsabs.harvard.edu/abs/2007ApJ...654...53M}{654, 53}

\bibitem[{Metuki {et~al.}(2016)Metuki, Libeskind, \& Hoffman}]{Metuki2016}
Metuki, O., Libeskind, N.~I., \& Hoffman, Y. 2016, \href{http://dx.doi.org/10.1093/mnras/stw979}{\color{magenta}\mnras}, 460

\bibitem[{Metuki {et~al.}(2015)Metuki, Libeskind, Hoffman, Crain, \& Theuns}]{Metuki2015}
Metuki, O., Libeskind, N.~I., Hoffman, Y., Crain, R.~A., \& Theuns, T. 2015, \href{http://dx.doi.org/10.1093/mnras/stu2166}{\color{magenta}\mnras}, 446

\bibitem[{{Miraghaei}(2020)}]{2020AJ....160..227M}
{Miraghaei}, H. 2020, \href{http://dx.doi.org/10.3847/1538-3881/abafb1}{\color{magenta}\aj}, \href{https://ui.adsabs.harvard.edu/abs/2020AJ....160..227M}{160, 227}

\bibitem[{Molin{\'e} {et~al.}(2022)Molin{\'e}, S{\'a}nchez-Conde, Aguirre-Santaella, Ishiyama, Prada, Cora, Croton, Jullo, Metcalf, Oogi, \& Ruedas}]{10.1093/mnras/stac2930}
Molin{\'e}, {\'A}., S{\'a}nchez-Conde, M.~A., Aguirre-Santaella, A., {et~al.} 2022, \href{http://dx.doi.org/10.1093/mnras/stac2930}{\color{magenta}\mnras}, 518, 157

\bibitem[{{Molin{\'e}} {et~al.}(2023){Molin{\'e}}, {S{\'a}nchez-Conde}, {Aguirre-Santaella}, {Ishiyama}, {Prada}, {Cora}, {Croton}, {Jullo}, {Metcalf}, {Oogi}, \& {Ruedas}}]{2023MNRAS.518..157M}
{Molin{\'e}}, {\'A}., {S{\'a}nchez-Conde}, M.~A., {Aguirre-Santaella}, A., {et~al.} 2023, \href{http://dx.doi.org/10.1093/mnras/stac2930}{\color{magenta}\mnras}, \href{https://ui.adsabs.harvard.edu/abs/2023MNRAS.518..157M}{518, 157}

\bibitem[{Muldrew {et~al.}(2011)Muldrew, Pearce, \& Power}]{10.1111/j.1365-2966.2010.17636.x}
Muldrew, S.~I., Pearce, F.~R., \& Power, C. 2011, \href{http://dx.doi.org/10.1111/j.1365-2966.2010.17636.x}{\color{magenta}\mnras}, 410, 2617

\bibitem[{{M{\"u}ller} {et~al.}(2018){M{\"u}ller}, {Pawlowski}, {Jerjen}, \& {Lelli}}]{Mueller2018}
{M{\"u}ller}, O., {Pawlowski}, M.~S., {Jerjen}, H., \& {Lelli}, F. 2018, \href{http://dx.doi.org/10.1126/science.aao1858}{\color{magenta}Science}, \href{https://ui.adsabs.harvard.edu/abs/2018Sci...359..534M}{359, 534}

\bibitem[{{Newton} {et~al.}(2021){Newton}, {Leo}, {Cautun}, {Jenkins}, {Frenk}, {Lovell}, {Helly}, {Benson}, \& {Cole}}]{Newton2021}
{Newton}, O., {Leo}, M., {Cautun}, M., {et~al.} 2021, \href{http://dx.doi.org/10.1088/1475-7516/2021/08/062}{\color{magenta}\jcap}, \href{https://ui.adsabs.harvard.edu/abs/2021JCAP...08..062N}{2021, 062}

\bibitem[{{Newton} {et~al.}(2024){Newton}, {Lovell}, {Frenk}, {Jenkins}, {Helly}, {Cole}, \& {Benson}}]{2024arXiv240816042N}
{Newton}, O., {Lovell}, M.~R., {Frenk}, C.~S., {et~al.} 2024, \href{https://ui.adsabs.harvard.edu/abs/2024arXiv240816042N}{\href{http://dx.doi.org/10.48550/arXiv.2408.16042}{\color{magenta}arXiv e-prints}, arXiv:2408.16042}

\bibitem[{{Nurmi} {et~al.}(2007){Nurmi}, {Hein{\"a}m{\"a}ki}, {Holopainen}, {Pihajoki}, {Saar}, {Einasto}, \& {Einasto}}]{2007IAUS..235..127N}
{Nurmi}, P., {Hein{\"a}m{\"a}ki}, P., {Holopainen}, J., {et~al.} 2007, in Galaxy Evolution across the Hubble Time, ed. F.~{Combes} \& J.~{Palou{\v{s}}}, Vol. 235, \href{https://ui.adsabs.harvard.edu/abs/2007IAUS..235..127N}{127--127}

\bibitem[{Onions {et~al.}(2012)Onions, Knebe, Pearce, Muldrew, Lux, Knollmann, Ascasibar, Behroozi, Elahi, Han, Maciejewski, Merchán, Neyrinck, Ruiz, Sgró, Springel, \& Tweed}]{10.1111/j.1365-2966.2012.20947.x}
Onions, J., Knebe, A., Pearce, F.~R., {et~al.} 2012, \href{http://dx.doi.org/10.1111/j.1365-2966.2012.20947.x}{\color{magenta}\mnras}, 423, 1200

\bibitem[{Patiri {et~al.}(2006)Patiri, Cuesta, Prada, Betancort-Rijo, \& Klypin}]{Patiri2006}
Patiri, S.~G., Cuesta, A.~J., Prada, F., Betancort-Rijo, J., \& Klypin, A. 2006, \href{http://dx.doi.org/10.1086/510330}{\color{magenta}\apj}, 652

\bibitem[{{Pawlowski} {et~al.}(2012{\natexlab{a}}){Pawlowski}, {Kroupa}, {Angus}, {de Boer}, {Famaey}, \& {Hensler}}]{PawlowskiFilamentary2012}
{Pawlowski}, M.~S., {Kroupa}, P., {Angus}, G., {et~al.} 2012{\natexlab{a}}, \href{http://dx.doi.org/10.1111/j.1365-2966.2012.21169.x}{\color{magenta}\mnras}, \href{https://ui.adsabs.harvard.edu/abs/2012MNRAS.424...80P}{424, 80}

\bibitem[{{Pawlowski} {et~al.}(2012{\natexlab{b}}){Pawlowski}, {Pflamm-Altenburg}, \& {Kroupa}}]{Pawlowski2012}
{Pawlowski}, M.~S., {Pflamm-Altenburg}, J., \& {Kroupa}, P. 2012{\natexlab{b}}, \href{http://dx.doi.org/10.1111/j.1365-2966.2012.20937.x}{\color{magenta}\mnras}, \href{https://ui.adsabs.harvard.edu/abs/2012MNRAS.423.1109P}{423, 1109}

\bibitem[{{Peebles}(1980)}]{1980lssu.book.....P}
{Peebles}, P.~J.~E. 1980, {The large-scale structure of the universe}

\bibitem[{{Pomar{\`e}de} {et~al.}(2017){Pomar{\`e}de}, {Hoffman}, {Courtois}, \& {Tully}}]{2017ApJ...845...55P}
{Pomar{\`e}de}, D., {Hoffman}, Y., {Courtois}, H.~M., \& {Tully}, R.~B. 2017, \href{http://dx.doi.org/10.3847/1538-4357/aa7f78}{\color{magenta}\apj}, \href{https://ui.adsabs.harvard.edu/abs/2017ApJ...845...55P}{845, 55}

\bibitem[{Reed {et~al.}(2005)Reed, Governato, Quinn, Gardner, Stadel, \& Lake}]{Reed2005}
Reed, D., Governato, F., Quinn, T., {et~al.} 2005, \href{http://dx.doi.org/10.1111/j.1365-2966.2005.09020.x}{\color{magenta}\mnras}, 359

\bibitem[{Rey {et~al.}(2019)Rey, Pontzen, \& Saintonge}]{10.1093/mnras/stz552}
Rey, M.~P., Pontzen, A., \& Saintonge, A. 2019, \href{http://dx.doi.org/10.1093/mnras/stz552}{\color{magenta}\mnras}, 485, 1906

\bibitem[{Sawala {et~al.}(2013)Sawala, Frenk, Crain, Jenkins, Schaye, Theuns, \& Zavala}]{10.1093/mnras/stt259}
Sawala, T., Frenk, C.~S., Crain, R.~A., {et~al.} 2013, \href{http://dx.doi.org/10.1093/mnras/stt259}{\color{magenta}\mnras}, 431, 1366

\bibitem[{{Shandarin} \& {Zeldovich}(1989)}]{Shandarin1989}
{Shandarin}, S.~F. \& {Zeldovich}, Y.~B. 1989, \href{http://dx.doi.org/10.1103/RevModPhys.61.185}{\color{magenta}Reviews of Modern Physics}, \href{https://ui.adsabs.harvard.edu/abs/1989RvMP...61..185S}{61, 185}

\bibitem[{Sheth \& Tormen(2004)}]{10.1111/j.1365-2966.2004.07733.x}
Sheth, R.~K. \& Tormen, G. 2004, \href{http://dx.doi.org/10.1111/j.1365-2966.2004.07733.x}{\color{magenta}\mnras}, 350, 1385

\bibitem[{Sousbie(2011)}]{10.1111/j.1365-2966.2011.18394.x}
Sousbie, T. 2011, \href{http://dx.doi.org/10.1111/j.1365-2966.2011.18394.x}{\color{magenta}\mnras}, 414, 350

\bibitem[{{Springel} {et~al.}(2008){Springel}, {Wang}, {Vogelsberger}, {Ludlow}, {Jenkins}, {Helmi}, {Navarro}, {Frenk}, \& {White}}]{2008MNRAS.391.1685S}
{Springel}, V., {Wang}, J., {Vogelsberger}, M., {et~al.} 2008, \href{http://dx.doi.org/10.1111/j.1365-2966.2008.14066.x}{\color{magenta}\mnras}, \href{https://ui.adsabs.harvard.edu/abs/2008MNRAS.391.1685S}{391, 1685}

\bibitem[{{Springel} {et~al.}(2005){Springel}, {White}, {Jenkins}, {Frenk}, {Yoshida}, {Gao}, {Navarro}, {Thacker}, {Croton}, {Helly}, {Peacock}, {Cole}, {Thomas}, {Couchman}, {Evrard}, {Colberg}, \& {Pearce}}]{2005Natur.435..629S}
{Springel}, V., {White}, S. D.~M., {Jenkins}, A., {et~al.} 2005, \href{http://dx.doi.org/10.1038/nature03597}{\color{magenta}\nat}, \href{https://ui.adsabs.harvard.edu/abs/2005Natur.435..629S}{435, 629}

\bibitem[{Springel {et~al.}(2001)Springel, White, Tormen, \& Kauffmann}]{10.1046/j.1365-8711.2001.04912.x}
Springel, V., White, S. D.~M., Tormen, G., \& Kauffmann, G. 2001, \href{http://dx.doi.org/10.1046/j.1365-8711.2001.04912.x}{\color{magenta}\mnras}, 328, 726

\bibitem[{Stoehr {et~al.}(2002)Stoehr, White, Tormen, \& Springel}]{10.1046/j.1365-8711.2002.05891.x}
Stoehr, F., White, S. D.~M., Tormen, G., \& Springel, V. 2002, \href{http://dx.doi.org/10.1046/j.1365-8711.2002.05891.x}{\color{magenta}\mnras}, 335, L84

\bibitem[{{Tegmark} {et~al.}(2004){Tegmark}, {Blanton}, {Strauss}, {Hoyle}, {Schlegel}, {Scoccimarro}, {Vogeley}, {Weinberg}, {Zehavi}, {Berlind}, {Budavari}, {Connolly}, {Eisenstein}, {Finkbeiner}, {Frieman}, {Gunn}, {Hamilton}, {Hui}, {Jain}, {Johnston}, {Kent}, {Lin}, {Nakajima}, {Nichol}, {Ostriker}, {Pope}, {Scranton}, {Seljak}, {Sheth}, {Stebbins}, {Szalay}, {Szapudi}, {Verde}, {Xu}, {Annis}, {Bahcall}, {Brinkmann}, {Burles}, {Castander}, {Csabai}, {Loveday}, {Doi}, {Fukugita}, {Gott}, {Hennessy}, {Hogg}, {Ivezi{\'c}}, {Knapp}, {Lamb}, {Lee}, {Lupton}, {McKay}, {Kunszt}, {Munn}, {O'Connell}, {Peoples}, {Pier}, {Richmond}, {Rockosi}, {Schneider}, {Stoughton}, {Tucker}, {Vanden Berk}, {Yanny}, {York}, \& {SDSS Collaboration}}]{2004ApJ...606..702T}
{Tegmark}, M., {Blanton}, M.~R., {Strauss}, M.~A., {et~al.} 2004, \href{http://dx.doi.org/10.1086/382125}{\color{magenta}\apj}, \href{https://ui.adsabs.harvard.edu/abs/2004ApJ...606..702T}{606, 702}

\bibitem[{Tempel {et~al.}(2015)Tempel, Guo, Kipper, \& Libeskind}]{10.1093/mnras/stv919}
Tempel, E., Guo, Q., Kipper, R., \& Libeskind, N.~I. 2015, \href{http://dx.doi.org/10.1093/mnras/stv919}{\color{magenta}\mnras}, 450, 2727

\bibitem[{{van de Weygaert} \& {Platen}(2011)}]{2011IJMPS...1...41V}
{van de Weygaert}, R. \& {Platen}, E. 2011, in International Journal of Modern Physics Conference Series, Vol.~1, International Journal of Modern Physics Conference Series, \href{https://ui.adsabs.harvard.edu/abs/2011IJMPS...1...41V}{41--66}

\bibitem[{{van den Bosch} \& {Jiang}(2016)}]{vdBoschJiang2016}
{van den Bosch}, F.~C. \& {Jiang}, F. 2016, \href{http://dx.doi.org/10.1093/mnras/stw440}{\color{magenta}\mnras}, \href{https://ui.adsabs.harvard.edu/abs/2016MNRAS.458.2870V}{458, 2870}

\bibitem[{{van den Bosch} \& {Ogiya}(2018)}]{vdBoschOgiya2018}
{van den Bosch}, F.~C. \& {Ogiya}, G. 2018, \href{http://dx.doi.org/10.1093/mnras/sty084}{\color{magenta}\mnras}, \href{https://ui.adsabs.harvard.edu/abs/2018MNRAS.475.4066V}{475, 4066}

\bibitem[{Van Den~Bosch {et~al.}(2005)Van Den~Bosch, Tormen, \& Giocoli}]{10.1111/j.1365-2966.2005.08964.x}
Van Den~Bosch, F.~C., Tormen, G., \& Giocoli, C. 2005, \href{http://dx.doi.org/10.1111/j.1365-2966.2005.08964.x}{\color{magenta}\mnras}, 359, 1029

\bibitem[{{Wang} {et~al.}(2023){Wang}, {Cooper}, {Bose}, {Frenk}, \& {Hellwing}}]{Wang2023}
{Wang}, C.-W., {Cooper}, A.~P., {Bose}, S., {Frenk}, C.~S., \& {Hellwing}, W.~A. 2023, \href{http://dx.doi.org/10.3847/1538-4357/ad011d}{\color{magenta}\apj}, \href{https://ui.adsabs.harvard.edu/abs/2023ApJ...958..166W}{958, 166}

\bibitem[{{Wang} {et~al.}(2007){Wang}, {Mo}, \& {Jing}}]{Wang2007}
{Wang}, H.~Y., {Mo}, H.~J., \& {Jing}, Y.~P. 2007, \href{http://dx.doi.org/10.1111/j.1365-2966.2006.11316.x}{\color{magenta}\mnras}, \href{https://ui.adsabs.harvard.edu/abs/2007MNRAS.375..633W}{375, 633}

\bibitem[{Wang {et~al.}(2017)Wang, Gonzalez-Perez, Xie, Cooper, Frenk, Gao, Hellwing, Helly, Lovell, \& Jiang}]{Wang2017}
Wang, L., Gonzalez-Perez, V., Xie, L., {et~al.} 2017, \href{http://dx.doi.org/10.1093/mnras/stx788}{\color{magenta}\mnras}, 468

\bibitem[{{Wang} {et~al.}(2020){Wang}, {Libeskind}, {Tempel}, {Pawlowski}, {Kang}, \& {Guo}}]{Wang2020}
{Wang}, P., {Libeskind}, N.~I., {Tempel}, E., {et~al.} 2020, \href{http://dx.doi.org/10.3847/1538-4357/aba6ea}{\color{magenta}\apj}, \href{https://ui.adsabs.harvard.edu/abs/2020ApJ...900..129W}{900, 129}

\bibitem[{{Wechsler} {et~al.}(2002){Wechsler}, {Bullock}, {Primack}, {Kravtsov}, \& {Dekel}}]{2002ApJ...568...52W}
{Wechsler}, R.~H., {Bullock}, J.~S., {Primack}, J.~R., {Kravtsov}, A.~V., \& {Dekel}, A. 2002, \href{http://dx.doi.org/10.1086/338765}{\color{magenta}\apj}, \href{https://ui.adsabs.harvard.edu/abs/2002ApJ...568...52W}{568, 52}

\bibitem[{Wechsler {et~al.}(2006)Wechsler, Zentner, Bullock, Kravtsov, \& Allgood}]{Wechsler_2006}
Wechsler, R.~H., Zentner, A.~R., Bullock, J.~S., Kravtsov, A.~V., \& Allgood, B. 2006, \href{http://dx.doi.org/10.1086/507120}{\color{magenta}\apj}, 652, 71

\bibitem[{White \& Frenk(1991)}]{White1991}
White, S. D.~M. \& Frenk, C.~S. 1991, \href{http://dx.doi.org/10.1086/170483}{\color{magenta}\apj}, 379

\bibitem[{White \& Rees(1978)}]{White1978}
White, S. D.~M. \& Rees, M.~J. 1978, \href{http://dx.doi.org/10.1093/mnras/183.3.341}{\color{magenta}\mnras}, 183

\bibitem[{Xu {et~al.}(2020)Xu, Guo, Zheng, Gao, Lacey, Gu, Liao, Shao, Mao, Zhang, \& Chen}]{Xu2020}
Xu, W., Guo, Q., Zheng, H., {et~al.} 2020, \href{http://dx.doi.org/10.1093/mnras/staa2497}{\color{magenta}\mnras}, 498

\bibitem[{{Zel'dovich}(1970)}]{Zeldovich1970}
{Zel'dovich}, Y.~B. 1970, \aap, \href{https://ui.adsabs.harvard.edu/abs/1970A&A.....5...84Z}{5, 84}

\bibitem[{Zhang {et~al.}(2021)Zhang, Yang, \& Guo}]{10.1093/mnras/stab2487}
Zhang, Y., Yang, X., \& Guo, H. 2021, \href{http://dx.doi.org/10.1093/mnras/stab2487}{\color{magenta}\mnras}, 507, 5320

\end{thebibliography}

\label{LastPage}
\end{document}